**Mathematical Discovery of Potential Therapeutic Targets: Application to Rare Melanomas**

Mahya Aghaee[1+], Victoria Cicchirillo[2+], Rowan Milner[2], Kyle Adams[3], Julia Bruner[4], William Hager[3], Ashley N. Brown[5], Elias Sayour[6], Domenico Santoro[2], Alberto Riva[7], Bently Doonan[8*], Helen Moore[1*]
+contributed equally; *contributed equally

1. Department of Medicine, College of Medicine, University of Florida, Gainesville, Florida, USA.
2. Small Animal Clinical Sciences, College of Veterinary Medicine, University of Florida, Gainesville, Florida, USA.
3. Department of Mathematics, College of Liberal Arts and Sciences, University of Florida, Gainesville, Florida, USA.
4. Department of Surgery, College of Medicine, University of Florida, Gainesville, Florida, USA.
5. Institute for Therapeutic Innovation, Department of Medicine, College of Medicine, University of Florida, Orlando, Florida, USA.
6. Department of Neurosurgery, College of Medicine, University of Florida, Gainesville, Florida, USA.
7. Fondazione Human Technocore, Milan, Italy.
8. Mayo Clinic, Jacksonville, FL USA.

# Abstract

Patients with rare types of melanoma such as acral, mucosal, or uveal melanoma, have lower survival rates than patients with cutaneous melanoma; these lower survival rates reflect the lower objective response rates to immunotherapy compared to cutaneous melanoma. Understanding tumor-immune dynamics in rare melanomas is critical for the development of new therapies and for improving response rates to current cancer therapies. Progress has been hindered by the lack of clinical data and the need for better preclinical models of rare melanomas. Canine melanoma provides a valuable comparative oncology model for rare types of human melanomas. We analyzed RNA sequencing data from canine melanoma patients and combined this with literature information to create a novel mechanistic mathematical model of melanoma-immune dynamics. Sensitivity analysis of the mathematical model indicated influential pathways in the dynamics, providing support for potential new therapeutic targets and future combinations of therapies. We share our learnings from this work, to help enable the application of this proof-of-concept workflow to other rare disease settings with sparse available data.

# Introduction

Melanoma is the most-lethal form of skin cancer, and is the 6th-most common cancer overall in the U.S., with over 100,000 new cases and 8,400 deaths expected in 2025.[1] The first immune checkpoint inhibitor (ICI) immunotherapy for cancer, the anti-CTLA-4 therapy ipilimumab, was approved in 2011 for patients with metastatic melanoma.[2] Before then, patients with metastatic melanoma had a median overall survival of approximately one year or less, regardless of the therapy used.[3] The median overall survival for metastatic melanoma patients treated with a combination of the anti-CTLA-4 and anti-PD-1 immunotherapies is now approximately 6 years, with melanoma-specific median survival of more than 10 years.[4] Recent approvals of anti-LAG-3 antibodies and tumor-infiltrating lymphocyte (TIL) therapies are poised to improve these numbers further.[5–7] However, a significant number of metastatic melanoma patients still do not respond to immunotherapy or experience relapse.[8] In particular, patients with the rare melanoma variants acral, mucosal, and uveal melanoma, experience significantly less benefit from immunotherapy or targeted therapy than patients with cutaneous melanoma.[9–12] Proposed mechanisms for this lack of response to immunotherapy in patients with rare melanoma variants include altered neoantigen presentation, lower somatic tumor mutational burden, more-suppressive tumor-immune microenvironments (TIME), and lower levels of TILs.[13] Strategies to overcome these mechanisms are urgently needed, along with an improved understanding of how and when to use immunotherapy to maximize therapeutic effect and ultimately improve patient outcomes.

With an expanding toolbox of therapies for melanoma, matching appropriate therapies with patients has become increasingly important. However, challenges remain in selecting treatments tailored to individual patients, identifying patient immune-specific needs, or altering/adjusting therapy based on tumor-immune interactions, beyond cases of disease progression or treatment-related toxicities.[14] An improved understanding of the TIME, and dynamics and response to therapy, could help with optimal treatment selection, optimization of regimens, and identification of new classes of therapies.[15]

Mechanistic mathematical models provide excellent tools for understanding and exploring disease and treatment dynamics. Such models have played an important role in medicine and the development of new therapies.[16,17] Mathematical models of tumor-immune dynamics are well-represented in the literature. Sachs et al. reviewed some of the simplest commonly-used models of tumor growth.[18] Eftimie et al. reviewed mechanistic models of tumor-immune dynamics of varying complexity, all of which consisted of ordinary differential equations (ODEs).[19] The books of Adam and Bellomo,[20] and Eladdadi et al.,[21] cover additional such tumor-immune models. Albrecht et al. reviewed computational models specific for melanoma.[22] Flach et al. modeled the dynamics of a melanoma tumor and fibroblasts.[23] An updated version of this model was later validated and extended to investigate drug resistance.[24] Kogan et al. modeled the dynamics among Th1 cells, Th2 cells, IL-10, and IFN-γ for melanoma.[25] Cappuccio et al. modeled the dynamics of IL-21 concentration, natural killer cells, CD8+ T cells, cytotoxic proteins, and tumor mass, to evaluate treatment strategies for melanoma.[26] Eftimie has multiple models in a melanoma context, including one that analyzed repolarization of macrophages in a melanoma setting,[27] and another that specifically described dynamics between M1 macrophages, M2 macrophages, Th1 cells, and Th2 cells.[28] De Pillis et al. modeled dendritic cell therapy for melanoma.[29] Ramaj and Zou modeled oncolytic virotherapy for melanoma and considered oxygen levels, uninfected tumor cells, and infected tumor cells.[30] Tsur et al. developed an ODE model of antigen-presenting cells, effector CD8+ tumor-infiltrating lymphocytes, and melanoma cells to predict patients' benefit from pembrolizumab, an anti-PD-1 immune-checkpoint inhibitor.[31] Lai and Friedman modeled melanoma with partial differential equations, and included several types of immune cells, multiple cytokines, immune-checkpoint mechanisms, and two therapies.[32] Nave and Sigron extended this work by developing explicit analytical functions and simulating the system with treatments including cobimetinib, atezolizumab, and vemurafenib.[33] Kronik et al. adapted a model they had developed for glioblastoma with immunotherapy;[34] they modeled tumor cells, cytotoxic T lymphocytes (CTLs), TGF-β, IFN-γ, and MHC I molecules with ODEs.[35] Beck et al. described intratumoral dynamics of transferred CTLs by modeling CTLs, IFN-γ, PDCD1, CD274, LAG-3, and HAVCR2/TIM-3 using ODEs.[36] Khalili and Vatankhah expanded a general cancer model with chemo-immunotherapy by de Pillis et al.[37] to include Tregs, and used a melanoma setting as an example to propose new approaches for modeling cancer.[38] Rodrigues et al. described dynamics of melanoma, tumor-associated macrophages, and CAR T cells in order to study CAR T-cell immunotherapy.[39] Anbari et al. focused on understanding uveal melanoma by building a mathematical model with four compartments: central blood, peripheral organs and tissues, the tumor, and tumor-draining lymph nodes with the goal of finding predictive biomarkers.[40]

In this work, we built on some of these previously-published ODE mathematical models of melanoma-immune dynamics. We developed a new mechanistic mathematical model focused solely on immune cells and their interactions with melanoma to help improve our understanding of the tumor-immune microenvironment. To better understand the tumor-immune dynamics, we did not include any therapies in the model; however, our future goal is to use our model to optimize treatment regimens for melanoma. We developed an initial model based on literature information, and included 15 immune cell types and their dynamics in the melanoma TIME. Although this model included most of the reported interactions between immune cells and melanoma cells, a model of that size presents challenges. A large model requires a large number of parameter values to be estimated from data, literature, or other means, and this can be challenging to do. And a large model can require additional run time for simulations, which are needed for sensitivity analysis and to ensure model robustness.[41,42] Thus we used data from canine melanoma patients in order to reduce our model while still capturing key interactions between immune cells and melanoma cells.

The use of canine patients (as well as other animal patients) as a comparative oncology model for human cancers has been expanding over recent decades,[43,44] particularly for the study of cancers that are rare in humans.[44] Unlike traditional inbred mouse and other laboratory animal studies, comparative oncology utilizes naturally-occurring cancers in animal patients. Studying these patients provides significant advantages over laboratory animal models: their disease recapitulates the natural initiation and development of a tumor and metastases, and it occurs in a host with an intact immune system that was present in the tumor's development and progression. Spontaneously-occurring canine melanoma cases are almost always non-cutaneous, with the most-common types being mucosal, acral, or uveal.[45] These types exhibit molecular and histopathological similarities to the corresponding types of human melanoma, but the higher prevalence in canines (compared to humans) enables detailed study of melanomas that are rare in human patients.[44,46]

We analyzed tissue samples from canine melanoma tumors and from healthy canines and obtained bulk RNA sequencing (RNA-seq) gene expression levels for each sample. We used immune cell deconvolution to determine levels of immune cell types in each sample. This deconvolution compares gene expression signatures of certain known immune cell types with bulk RNA-seq gene expression data, and quantifies the relative abundance of these cells within a sample that contains a mixture of different cell types. Cell types with statistically significant differences between tumor and healthy samples were considered as candidates for key immune cells to include in our final melanoma tumor-immune model.

After finalizing the reduced model and estimating parameter values from the literature, we applied global sensitivity analysis to determine the most-influential parameters in the model. These parameters represent potential novel targets and could also be used to inform the selection of therapies to use in combination therapy. We applied identifiability analysis to find a subset of these influential parameters that could also be estimated from the expected available data.

# Materials and Methods

## Data

We collected and analyzed tissue samples from canine melanoma patients with a mix of these rare melanomas, in order to focus on the rare and non-responsive melanoma types. The samples we analyzed included 8 primary melanoma tumor samples, 4 healthy skin samples, 8 metastatic cancerous lymph node (LN) tumor samples, and 3 healthy LN samples. Despite the fact that the LN is a proximal site for mounting immune defense against local tumors, the LN is the most-common site for early tumor metastasis.[47,48]

## Samples and Collection

Flash-frozen paraffin-embedded (FFPE) tumor samples were obtained from canine patients (n = 8) diagnosed with melanoma from 2016 to 2022 that were surgically removed at the University of Florida Small Animal Hospital. FFPE samples were also obtained from each animal from lymph nodes removed at the time of melanoma tumor removal. Patients were excluded if they were not naive to any form of therapeutic intervention including chemotherapy, radiation therapy, or immunotherapy. Samples were excluded if they did not have a histologically-confirmed metastatic lymph node removed at the same time as primary tumor removal. One healthy lymph node sample that did not meet quality control standards as specified and reviewed by the nanoString Information Technologies team (nanoString, Bothell, WA, USA) for gene expression analysis was also excluded. See Table 1 for more information about the samples and animals they were taken from.[49]

| Sample Label | Patient ID | Tumor Type | Sample Type | Sample Location | Breed | Age (yrs) | Sex |
|---|---|---|---|---|---|---|---|
| T 1 | 1 | Primary melanoma | FFPE | Left mandible | Mixed | 14 | Fs |

| | | (mucosal) | | | | | |
|---|---|---|---|---|---|---|---|
| T 2 | 2 | Primary melanoma (acral) | FFPE | Left front 4th digit | Doberman Pinscher | 8 | Fs |
| T 3 | 3 | Primary melanoma (acral) | FFPE | Left hind 3rd digit | Labrador Retriever | 10 | Mc |
| T 4 | 4 | Primary melanoma (mucosal) | FFPE | Tonsil | Labrador Retriever | 12 | Fs |
| T 5 | 5 | Primary melanoma (mucosal) | FFPE | Tonsil | Cocker Spaniel | 12 | Mc |
| T 6 | 6 | Primary melanoma (mucosal) | FFPE | Anal sac | Brussels Griffon | 11 | Mc |
| T 7 | 7 | Primary melanoma (acral) | FFPE | Right hind 4th digit | Golden Retriever | 9 | Mc |
| T 8 | 8 | Primary melanoma (mucosal) | FFPE | Right maxilla | Cocker Spaniel | 11 | Fs |
| LNT 1 | 1 | Metastatic melanoma | FFPE | Left submandibular lymph node | Mixed | 14 | Fs |
| LNT 2 | 2 | Metastatic melanoma | FFPE | Left prescapular lymph node | Doberman Pinscher | 8 | Fs |
| LNT 3 | 3 | Metastatic melanoma | FFPE | Left popliteal lymph node | Labrador Retriever | 10 | Mc |
| LNT 4 | 4 | Metastatic melanoma | FFPE | Left medial retropharyngeal lymph node | Labrador Retriever | 12 | Fs |
| LNT 5 | 5 | Metastatic melanoma | FFPE | Left retropharyngeal lymph node | Cocker Spaniel | 12 | Mc |

| LNT 6 | 6 | Metastatic melanoma | FFPE | Abdominal lymph node | Brussels Griffon | 11 | Mc |
|---|---|---|---|---|---|---|---|
| LNT 7 | 7 | Metastatic melanoma | FFPE | Right popliteal lymph node | Golden Retriever | 9 | Mc |
| LNT 8 | 8 | Metastatic melanoma | FFPE | Right submandibular lymph node | Cocker Spaniel | 11 | Fs |
| LNH 1 | 4 | Normal lymph node | FFPE | Left submandibular lymph node | Labrador Retriever | 12 | Fs |
| LNH 2 | 7 | Normal lymph node | FFPE | Right inguinal lymph node | Golden Retriever | 9 | Mc |
| LNH 3 | 8 | Normal lymph node | FFPE | Left retropharyngeal lymph node | Cocker Spaniel | 11 | Fs |

**Table 1. Patient information for primary and metastatic melanoma samples, and healthy lymph node samples.** T = primary tumor; LNT = cancerous lymph node tumor, metastasized from the primary tumor of the patient with the same patient identification number; LNH = healthy lymph node, from the patient with the specified identification number; ID = identification number; Fs = Female spayed; Mc = Male castrated. Patient IDs are unique and indicate which samples are from the same animals.

Healthy samples were epidermal explants from the abdomens of healthy dogs. These were obtained through IACUC # 201810437 protocol. For each sample, an 8mm biopsy was taken and cut in quarters. Then the skin was placed in 1.25 U/ml of dispase (Dispase I, Sigma-Aldrich, St. Louis, MO, USA) solution at 4°C overnight in sterile tubes. The next day the sample was placed in a sterile Petri dish with CnT-09® medium and the epidermis detached from the dermis. The epidermis was then subjected to extraction. See Table 2 for more information about the healthy samples and animals they were taken from.

| Label | Sample Type | Sample Location | Breed | Age | Sex |
|---|---|---|---|---|---|
| H 1 | Fresh | Abdomen | Mixed | 2y 7m | Fs |
| H 2 | Fresh | Abdomen | American Staffordshire terrier | 4y 3m | Mc |
| H 3 | Fresh | Abdomen | Mixed | 6y | Mc |

| H 4 | Fresh | Abdomen | Mixed | 2y 10m | Mc |
|---|---|---|---|---|---|

**Table 2. Information for healthy tissue samples from four healthy canines.** H = healthy sample; y = years; m = months; Fs = Female spayed; Mc = Male castrated.

## RNA Isolation

One 4 µm-thick section was cut from each FFPE block and discarded prior to cutting six 5-µm thick sections to use for RNA extraction. A new microblade was used in between cutting sections from each FFPE block. RNA extraction from the FFPE samples was performed using the RNeasy® FFPE kit (Qiagen, Valencia, CA) with an adjustment made to the manufacturer's protocol. After adding Proteinase K to the samples, all samples were incubated at 80 degrees Celsius for 60 minutes instead of 15 minutes to allow for more-thorough digestion of the tissue. The samples were purified using the RNeasy clean up (Qiagen, Valencia, CA, USA) per the manufacturers' protocols at the author's discretion. RNA quantity and quality (260/280 ratio) was measured using a Nanodrop Spectrophotometer (Thermo Fisher Scientific, Waltham, MA). The 2100 Bioanalyzer (Agilent Technologies, Santa Clara, CA) at the University of Florida's Interdisciplinary Center for Biotechnology Research was used to calculate the DV200 of each sample. Samples were included in the nCounter analysis if the total RNA concentration in the well was at least 100 ng. If 5µl of the sample contained >1000 ng of RNA, the sample was further diluted with RNase free water. Samples were defined as sufficient quality for analysis if the 260/280 ratio was ~2.1 and the DV200 was at least 50. All RNA samples were eluted with RNase free water and stored at -40 degrees Celsius after extraction prior to analysis.

## nanoString nCounter® Bulk RNA-seq Analysis

RNA samples underwent code set hybridization (nanoString, Bothell, WA) to the Canine IO probe and spike-in genes per the manufacturer's protocol. Samples were loaded into the nCounter MAX (nanoString, Bothell, WA) prep station and digital analyzer 16-18 hours later per the manufacturer's protocol. RNA bound to probes was counted with the nCounter® Digital Analyzer. The nCounter® data were analyzed on the proprietary software, nSolver™ (nanoString, Bothell, WA, USA). Data quality control was performed per the manufacturer's protocol. Raw gene expression data were normalized using a positive control normalization factor to account for variations in samples, lanes, cartridges, user technique, hybridization, complex-to-slide binding, and imaging. This factor was calculated using positive controls in every sample. A CodeSet content normalization factor was also used to remove input variance and account for different degradation states of the samples. This factor was calculated using the geNorm function26 to find the most-stable housekeeping genes (TLK2, TBP, ABCF1, NRDE2, SF3A1, ERCC3, PUM1, STK11IP, GUSB, DNAJC14, TBC1D10B, OAZ1, PORR2A, UBB, MRPL19, TMUB2, PSMC4, SDHA, TFRC, G6PD) via pairwise variance analysis. The software calculated the log2 geometric mean (fold change) of different sample groups and used t-tests to determine statistical significance. The Benjamini-Yekutieli false discovery rate method was utilized for p-value adjustments (adj p-value) to account for the expectation that significant changes in genes may be correlated with or dependent on each other, and an adjusted p-value of < 0.05 was used for statistical significance.[50]

## Immune cell deconvolution

Since tumor microenvironments are made up of multiple cell types, and not just malignant tumor cells, quantifying the various immune cells that infiltrate tumors is important for understanding the immune response to malignancy. Immune cell deconvolution uses reference gene expression signatures and statistical methods to quantitatively estimate the levels of various immune cells in a given tissue sample.[51] It separates the sources of RNA-seq data variations, such as batch effects of cell type-specific gene expression. We used deconvolution techniques to estimate the levels of immune cells in our data sets. Cell types that showed statistically significant differences between tumor samples and healthy tissue were considered important in tumor-immune dynamics and were retained in our mathematical model. We performed immune cell deconvolution on our bulk RNA-seq data using the original CIBERSORT method within the CIBERSORTx deconvolution algorithm (https://cibersortx.stanford.edu).[52] The LM22 signature matrix was used as the immune cell reference. CIBERSORTx was run using both relative fraction and absolute fraction output modes with quantile normalization disabled; the deconvolution was estimated with 100 permutations. This method has been used previously to accurately describe the immune cell response and expression specifically in melanoma.[53]

To compare primary tumor samples and healthy samples, and lymph node tumor samples and lymph node healthy samples, we used the Mann-Whitney U test to compare the central tendencies, since our data are continuous and non-normally distributed. Note that our primary melanoma and healthy sample data come from different subjects, and thus the data in different groups are independent. The Wilcoxon matched-pairs signed-rank test was used for the comparison of primary tumor samples and lymph node tumor samples, since they are matched samples from the same patients.

## Mathematical Model

To develop our mechanistic model of the tumor-immune dynamics, we started by reviewing the relevant literature. **Fig A1** in appendix A shows the initial model we developed, based on literature, which includes 15 immune cell types and melanoma. Detailed descriptions of the pathways are in **Table A1**. However, estimating all of the parameters in an initial model of this size is challenging, and the model size raises the computational cost of analysis compared to smaller models. Therefore, we used immune cell deconvolution as described above to reduce the model to a simpler one that captures key biological dynamics. Additional variables and parameters can be incorporated in future steps.

**Sensitivity Analysis**: Global sensitivity analysis (GSA) is a method that determines which parameters are most influential on the outcome or quantity of interest (QOI). These are the parameters that can contribute most to tumor regression if we change their values.[41] This information can be used to confirm some of the expected model behavior, look for potential novel targets, and decide which therapies to combine for maximal benefit. Optimal control techniques can then be used to determine regimens that maximize efficacy while simultaneously minimizing toxicity.[54–58]

Our QOI for this work was the number of melanoma cells at 180 days. All simulations were performed in MATLAB R2024b and we used ode15s for integrating the differential equations. We compared two different global sensitivity analysis approaches, Sobol and Extended Fourier

Amplitude Sensitivity Test (eFAST). We assumed each parameter's values were uniformly distributed with a range that spanned from 0.5 times its nominal value up to 1.5 times its nominal value. For sensitivity analysis, the choice of sampling scheme impacts convergence.[59] In the Sobol method, we used Sobol sequences for sampling, which improved the convergence speed and is less computationally expensive than Latin Hypercube Sampling. Moreover, Sobol sequences result in samples that are almost uniformly distributed over the parameter space. For eFAST, we use the standard eFAST sampling, which chooses points on an approximately space-filling curve with selected frequencies for the curve and the points.

**Structural Identifiability**: We performed structural identifiability to explore the possibility of estimating the influential parameters uniquely from data. Using sensitivity analysis allows us to fix all parameters except the most-influential ones without expecting major changes in the outcome of interest. Structural identifiability allows us to study the outcome's properties such as global and local identifiability, as well as non-identifiability. We used GenSSI,[60] an open-source software in the MATLAB environment, which uses Lie derivatives of the system to generate new equations, which are tested for independence. As emphasized by Sher et al., as well as Kao and Eisenberg,[61] testing for structural identifiability is important due to the possibility of misleading results of non-identifiable parameters.[62]

**Kolmogorov-Smirnov Test:** To test whether the most-influential parameters adequately capture the variability of the QOI, we used the two-sample Kolmogorov-Smirnov (KS) test. The first sample is the distribution of the QOI values obtained from simulating the model with all parameters varying. The second sample is the distribution of the QOI values obtained from simulating the model with only the most-influential parameters varying.

### RNA-seq analysis with iDEP

First described in publications in 2008, RNA-sequencing (RNA-seq) analysis is a technique to analyze gene expression levels of different cells.[63] We compared gene expression levels in canine patient primary tumor samples with those in healthy canine tissue samples. Gene expression read counts were imported into the R language-based integrated differential expression and pathway analysis package (iDEP, version .96, http://bioinformatics.sdstate.edu/idep96/).[64] The iDEP web application was used for data pre-processing and several analyses that we describe in this section. We extracted sufficient quality and quantity RNA for analysis of the included samples. We used DESeq2 in iDEP for differential gene expression analysis. Differentially-expressed genes (DEGs) were defined as those with a $|\text{fold-change}| > 2$, with adjusted p-values $< 0.05$, where the Benjamini-Hochberg method in iDEP was used to obtain p-values adjusted for multiple testing.[50] We ranked these using standard deviation across all samples. This ranking allowed us to establish a hierarchy of clusters based on gene expression patterns. We created a heat map to display these results. The default settings in iDEP were used for generating the heat map, which included using "correlation" distance metric and "average" linkage method. The selected DEGs were used to further investigate Gene Ontology (GO) functions and Kyoto Encyclopedia of Genes and Genomes (KEGG) pathways.

## Results

## Differential Gene Expression

Our main data analysis in this work compares primary canine melanoma tumors to healthy canine tissue samples. Principal component analysis (PCA) revealed a distinct difference between these groups, with the tumors clustering together distinctly from the healthy samples (**Fig 31B**). As seen in the volcano plot (**Fig 13C**), there were 367 upregulated genes and 148 downregulated genes.

## Immune Cell Deconvolution

We used our bulk RNA-seq as input of CIBERSORTx and LM22 as the signature matrix. The expression and abundance of subpopulation of immune cells was evaluated using CIBERSORTx. Our immune cell deconvolution found that regulatory T cells (p = 0.002), M1 macrophages (p = 0.004), M2 macrophages (p = 0.004), and CD8+ T cells (p = 0.004) are among immune cells that showed significant difference between tumor samples and healthy samples. These results align with external evidence reported for melanoma,[65] which found that CD8+ T cells, M2 macrophages, and regulatory T cells were among the top immune cells showing significant differences.

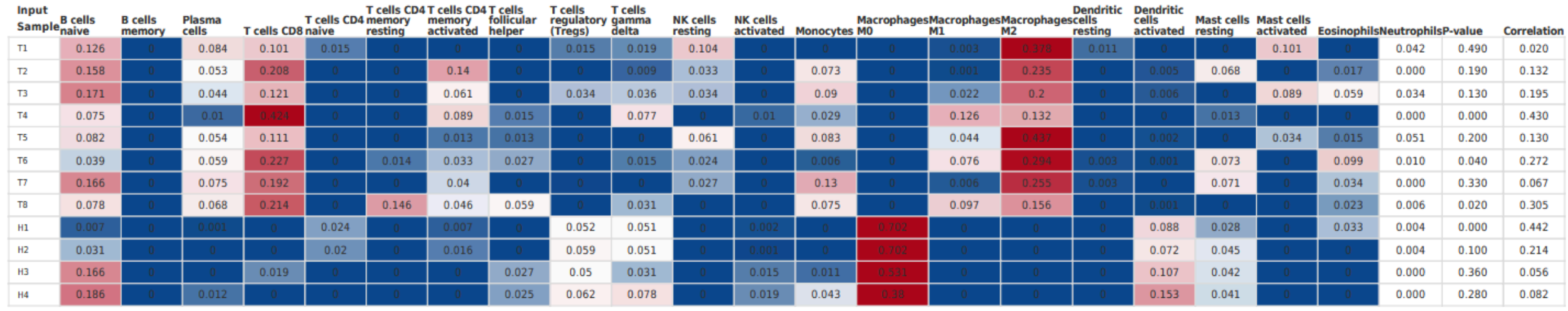

| Input Sample | B cells naive | B cells memory | Plasma cells | T cells CD8 | T cells CD4 naive | T cells CD4 memory resting | T cells CD4 memory activated | T cells follicular helper | T cells regulatory (Tregs) | T cells gamma delta | NK cells resting | NK cells activated | Monocytes | Macrophages M0 | Macrophages M1 | Macrophages M2 | Dendritic cells resting | Dendritic cells activated | Mast cells resting | Mast cells activated | Eosinophils | Neutrophils | P-value | Correlation |
|---|---|---|---|---|---|---|---|---|---|---|---|---|---|---|---|---|---|---|---|---|---|---|---|
| T1 | 0.126 | 0 | 0.084 | 0.101 | 0.015 | 0 | 0 | 0 | 0.015 | 0.019 | 0.104 | 0 | 0 | 0 | 0.003 | 0.378 | 0.011 | 0 | 0 | 0.101 | 0 | 0.042 | 0.490 | 0.020 |
| T2 | 0.158 | 0 | 0.053 | 0.208 | 0 | 0 | 0.14 | 0 | 0 | 0.009 | 0.033 | 0 | 0.073 | 0 | 0.001 | 0.235 | 0 | 0.005 | 0.068 | 0 | 0.017 | 0.000 | 0.190 | 0.132 |
| T3 | 0.171 | 0 | 0.044 | 0.121 | 0 | 0 | 0.061 | 0 | 0.034 | 0.036 | 0.034 | 0 | 0.09 | 0 | 0.022 | 0.2 | 0 | 0.006 | 0 | 0.089 | 0.059 | 0.034 | 0.130 | 0.195 |
| T4 | 0.075 | 0 | 0.01 | 0.424 | 0 | 0 | 0.089 | 0.015 | 0 | 0.077 | 0 | 0.01 | 0.029 | 0 | 0.126 | 0.132 | 0 | 0 | 0.013 | 0 | 0 | 0.000 | 0.000 | 0.430 |
| T5 | 0.082 | 0 | 0.054 | 0.111 | 0 | 0 | 0.013 | 0.013 | 0 | 0 | 0.061 | 0 | 0.083 | 0 | 0.044 | 0.437 | 0 | 0.002 | 0 | 0.034 | 0.015 | 0.051 | 0.200 | 0.130 |
| T6 | 0.039 | 0 | 0.059 | 0.227 | 0 | 0.014 | 0.033 | 0.027 | 0 | 0.015 | 0.024 | 0 | 0.006 | 0 | 0.076 | 0.294 | 0.003 | 0.001 | 0.073 | 0 | 0.099 | 0.010 | 0.040 | 0.272 |
| T7 | 0.166 | 0 | 0.075 | 0.192 | 0 | 0 | 0.04 | 0 | 0 | 0 | 0.027 | 0 | 0.13 | 0 | 0.006 | 0.255 | 0.003 | 0 | 0.071 | 0 | 0.034 | 0.000 | 0.330 | 0.067 |
| T8 | 0.078 | 0 | 0.068 | 0.214 | 0 | 0.146 | 0.046 | 0.059 | 0 | 0.031 | 0 | 0 | 0.075 | 0 | 0.097 | 0.156 | 0 | 0.001 | 0 | 0 | 0.023 | 0.006 | 0.020 | 0.305 |
| H1 | 0.007 | 0 | 0.001 | 0 | 0.024 | 0 | 0.007 | 0 | 0.052 | 0.051 | 0 | 0.002 | 0 | 0.702 | 0 | 0 | 0 | 0.088 | 0.028 | 0 | 0.033 | 0.004 | 0.000 | 0.442 |
| H2 | 0.031 | 0 | 0 | 0 | 0.02 | 0 | 0.016 | 0 | 0.059 | 0.051 | 0 | 0.001 | 0 | 0.702 | 0 | 0 | 0 | 0.072 | 0.045 | 0 | 0 | 0.004 | 0.100 | 0.214 |
| H3 | 0.166 | 0 | 0 | 0.019 | 0 | 0 | 0 | 0.027 | 0.05 | 0.031 | 0 | 0.015 | 0.011 | 0.531 | 0 | 0 | 0 | 0.107 | 0.042 | 0 | 0 | 0.000 | 0.360 | 0.056 |
| H4 | 0.186 | 0 | 0.012 | 0 | 0 | 0 | 0 | 0.025 | 0.062 | 0.078 | 0 | 0.019 | 0.043 | 0.38 | 0 | 0 | 0 | 0.153 | 0.041 | 0 | 0 | 0.000 | 0.280 | 0.082 |

**Fig 2. Heat map of estimated proportions of 22 immune cells for twelve canine samples (8 primary melanoma samples, 4 samples from healthy animals).** CD8+ T cells, M1 macrophages, M2 macrophages, and regulatory T cells show significant differences between groups. We used CIBERSORTx to generate this figure.

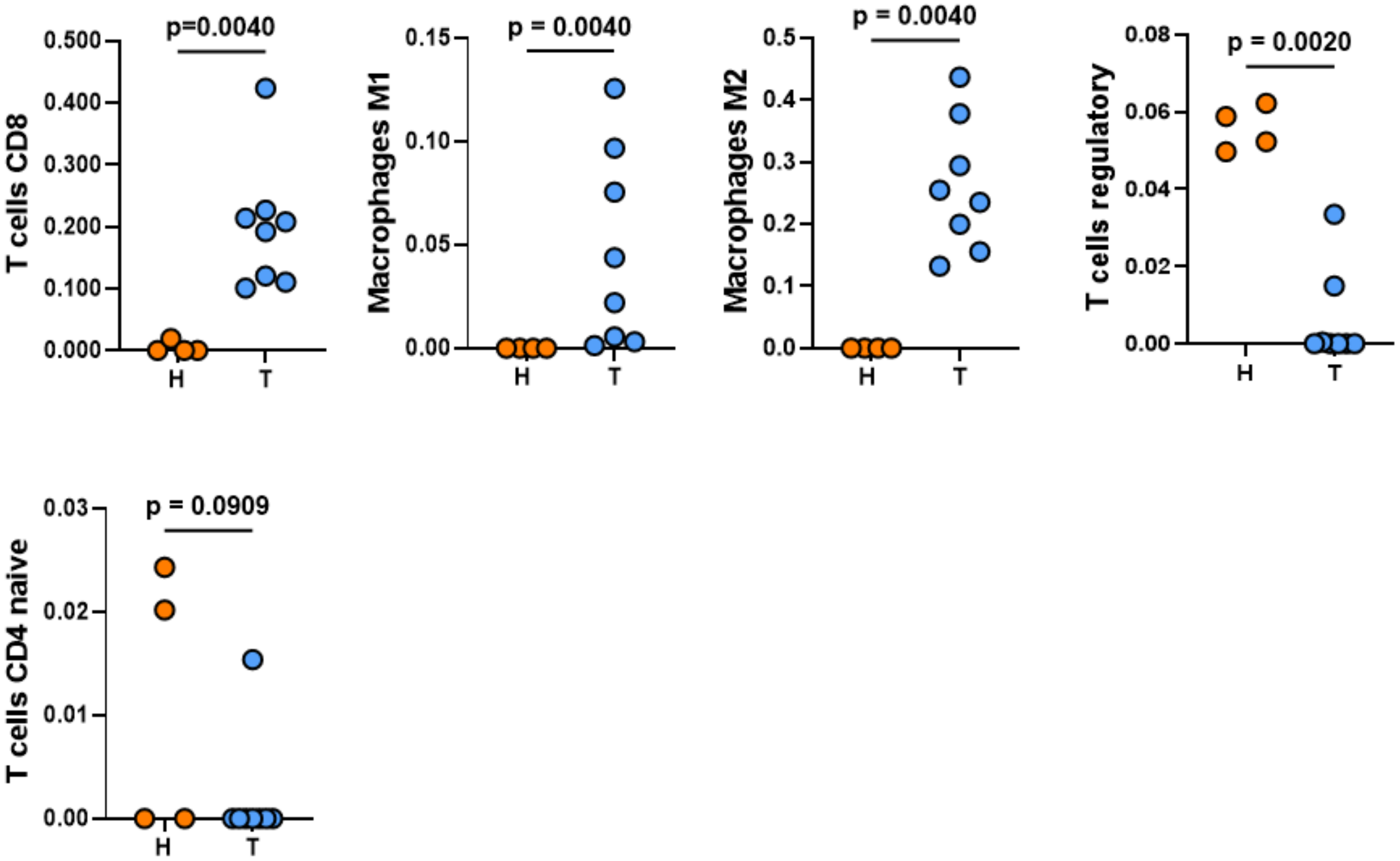

**Fig 3. Comparison of immune cell deconvolution cell counts for healthy vs tumor samples.** Analysis was conducted using CIBERSORTx. The four immune cell types on the top row, namely, CD8+ T cells, M1 and M2 macrophages, and regulatory T cells show significant differences between the primary tumor samples and healthy control samples. The CD4+ T cells on the bottom row are shown for comparison and do not show significant differences. All p-values were obtained using the Mann-Whitney U test in Prism. Graphing and pairwise statistical analyses were conducted in GraphPad Prism 10.

## Gene Ontology Enrichment Analysis

We performed Gene Ontology (GO) analysis to explore functional enrichment of DEGs in three categories - biological processes (BPs), cellular components (CCs), and molecular function (MFs). We ranked the results (**Table 3**) based on the number of genes in each term. The results show significant changes in BPs of upregulated DEGs mainly on immune-related responses such as immune system process, positive regulation of biological process, immune response, response to external stimulus, and response to stress. For CCs, upregulated DEGs were significantly enriched in cell periphery, plasma membrane, extracellular region, cell surface, and extracellular space, while significant changes in MFs from upregulated DEGs were in receptor related pathways such as signaling receptor binding, signaling receptor activity, molecular transducer activity, transmembrane signaling receptor activity, and cytokine receptor binding. Moreover, the analysis shows the downregulated genes were significantly enriched in BPs, CCs, and MFs.

Downregulated DEGs were mainly enriched for BPs in signaling, positive regulation of biological process, immune response, response to external stimulus, and response to stress, while cell periphery, plasma membrane, cytosol, cell surface, and intrinsic component of plasma membrane were enriched in CC, and identical protein binding, enzyme binding, signaling receptor binding, kinase activity, and phosphotransferase activity, alcohol group as acceptor were enriched in MF. **Fig 1C** shows the volcano plot of significant DEGs between primary tumor samples and healthy canine samples.

## Kyoto Encyclopedia of Genes and Genomes (KEGG) Analysis

KEGG is a tool that looks at gene signaling pathways, and thereby associates functional with genomic data. As observed in **Fig 4**, PI3K was highly upregulated in primary melanoma tumor samples; this finding aligns with human studies that showed PI3K is one of the most-important effector pathways in melanoma.[66,67] Moreover, a fundamental cell-cell adhesion protein called E-cadherin in our data; notably E-cadherin has been shown to be downregulated in metastatic human melanoma.[68] This loss of expression is believed to be a trait in the epithelial-mesenchymal transition of tumor metastasis.[68] Moreover, downregulation of E-cadherin is associated with potential metastasis in melanoma,[69] which we expected due the nature of our primary samples. The PTEN pathway was upregulated, which was not expected. Additionally, NRAS mutation was expected as it is known as second-most common mutation in metastatic melanoma in humans.[70] The result follows a very classical melanoma picture of upregulation in RAS pathway activity with NRAS and PI3K upregulation; the RAF being downregulated in these particular samples is likely just because of the abhorrent signal through NRAS in this sample size going more through P13K. Our data also align with human melanoma classical features of P53 loss and loss of cell cycle control with cyclin D2, which supports a classical melanoma picture in the canine.

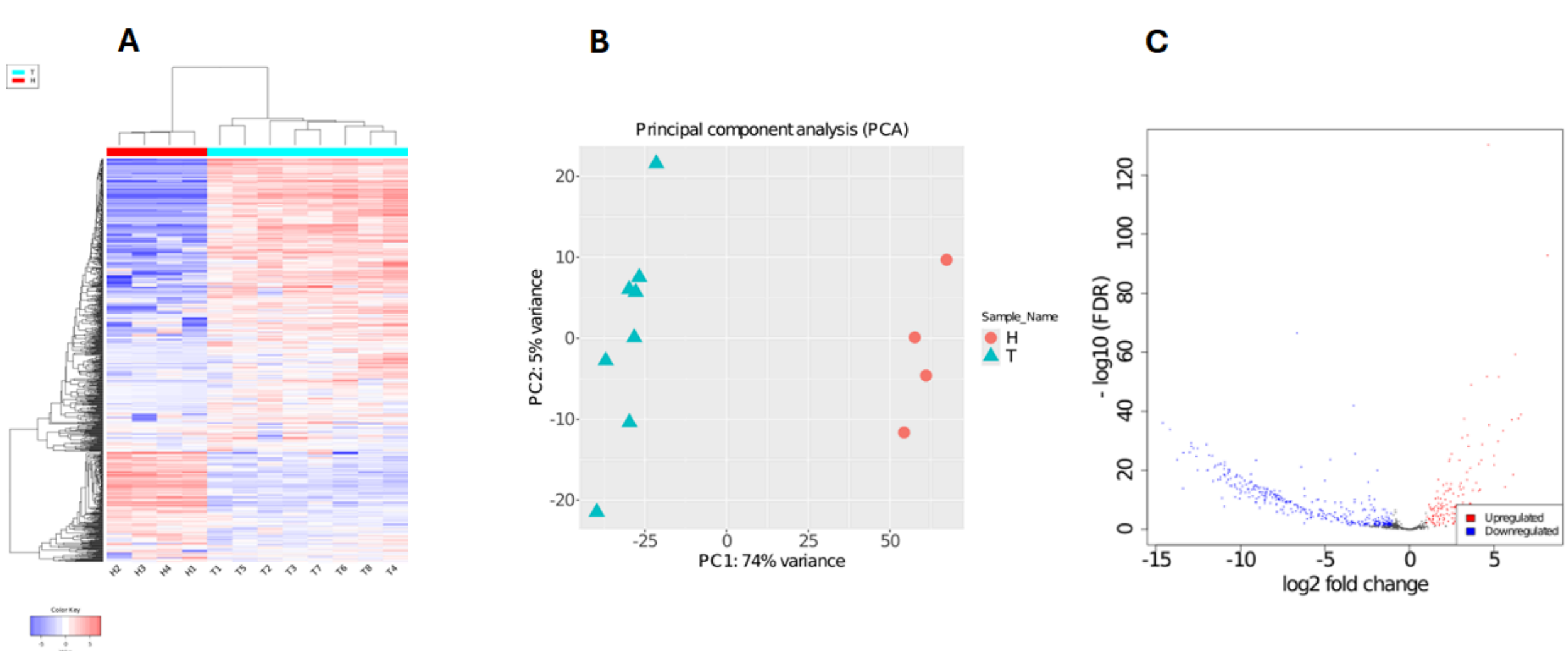

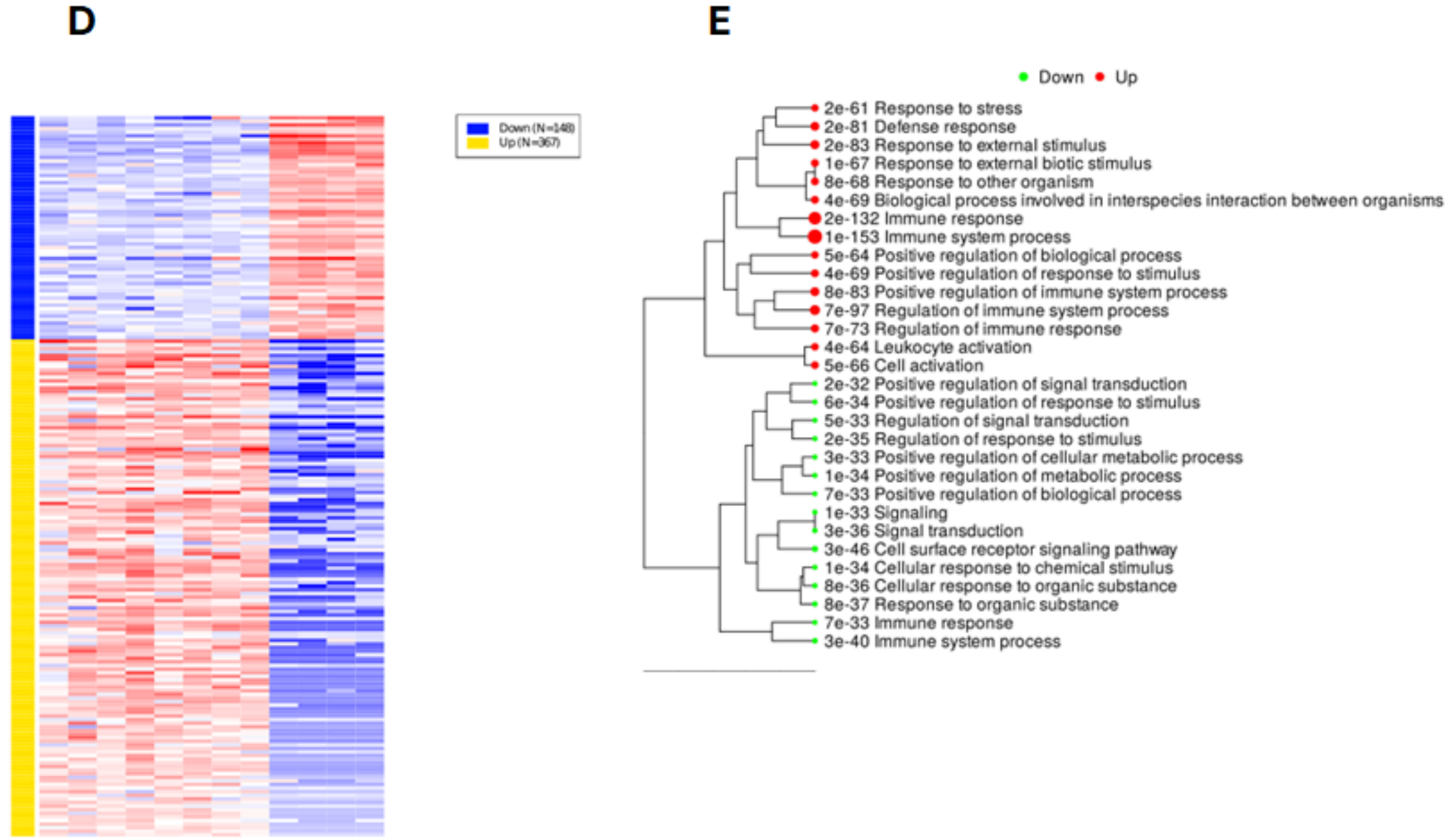

**Fig 3. Gene expression analysis.** (A) Hierarchical clustering heat map. The red bar on top indicates the columns for healthy samples and the turquoise bar indicates which ones are tumor samples. (B) Principal component analysis of the samples. PC1 was able to explain 74% of the variance of the data and PC2 was able to explain 5%. (C) Volcano plot showing how significantly differentiated the genes from Figure A are. (D) Heat map of downregulated (blue) and upregulated (yellow) genes and shows how similar the gene expression levels are with the group of tumor samples, as well as with the group of healthy samples. (E) Labels for the groups of genes shown in plot D. The p-values in plot E indicate how significant that group is. All plots in Fig. 3 were produced in iDEP version .96, and they represent analysis for primary melanoma tumor samples (T) vs healthy samples (H).

| Category | Term | Count | p-value |
|---|---|---|---|
| BP | Immune system process | 216 | 9.5e-147 |
| BP | Positive regulation of biological process | 211 | 4.7e-62 |
| BP | Immune response | 172 | 2.6e-127 |
| BP | Response to external stimulus | 159 | 6.3e-83 |
| BP | Response to stress | 158 | 5.9e-58 |
| CC | Cell periphery | 151 | 4.0e-22 |
| CC | Plasma membrane | 143 | 5.4e-22 |
| CC | Extracellular region | 102 | 7.6e-36 |

| CC | Cell surface | 85 | 4.4e-56 |
|---|---|---|---|
| CC | Extracellular space | 79 | 4.8e-32 |
| MF | Signaling receptor binding | 104 | 3.5e-52 |
| MF | Signaling receptor activity | 83 | 6.8e-21 |
| MF | Molecular transducer activity | 83 | 6.8e-21 |
| MF | Transmembrane signaling receptor activity | 69 | 3.4e-15 |
| MF | Cytokine receptor binding | 53 | 2.7e-47 |

**Table 3. The top 5 enriched GO terms for each category.** Upregulated pathways from Biological Processes (BPs), Cell Components (CCs), and Molecular Functions (MFs) are shown. iDEP version .96 was used for all analyses.

| Category | Term | Count | p-value |
|---|---|---|---|
| BP | Signaling | 104 | 1.0e-32 |
| BP | Positive regulation of biological process | 96 | 1.0e-31 |
| BP | Cell surface receptor signaling pathway | 83 | 1.7e-46 |
| BP | Regulation of response to stimulus | 80 | 1.1e-33 |
| BP | Positive regulation of metabolic process | 80 | 5.2e-33 |
| CC | Cell periphery | 63 | 5.5e-09 |
| CC | Plasma membrane | 60 | 5.2e-09 |
| CC | Cytosol | 41 | 8.4e-06 |
| CC | Cell surface | 29 | 1.8e-15 |
| CC | Intrinsic component of plasma membrane | 26 | 3.1e-08 |
| MF | Identical protein binding | 47 | 1.3e-19 |
| MF | Enzyme binding | 45 | 8.8e-17 |
| MF | Signaling receptor binding | 36 | 1.4e-14 |
| MF | Kinase activity | 25 | 3.6e-10 |

| MF | Phosphotransferase activity, alcohol group as acceptor | 25 | 5.4e-11 |
|---|---|---|---|

**Table 4. Enriched GO terms**. The top 5 enriched GO terms with downregulated pathways from Biological Process (BP), Cell Components (CC), and Molecular Functions (MF) are selected. All analyses (BP) - (MF) used iDEP version .96.

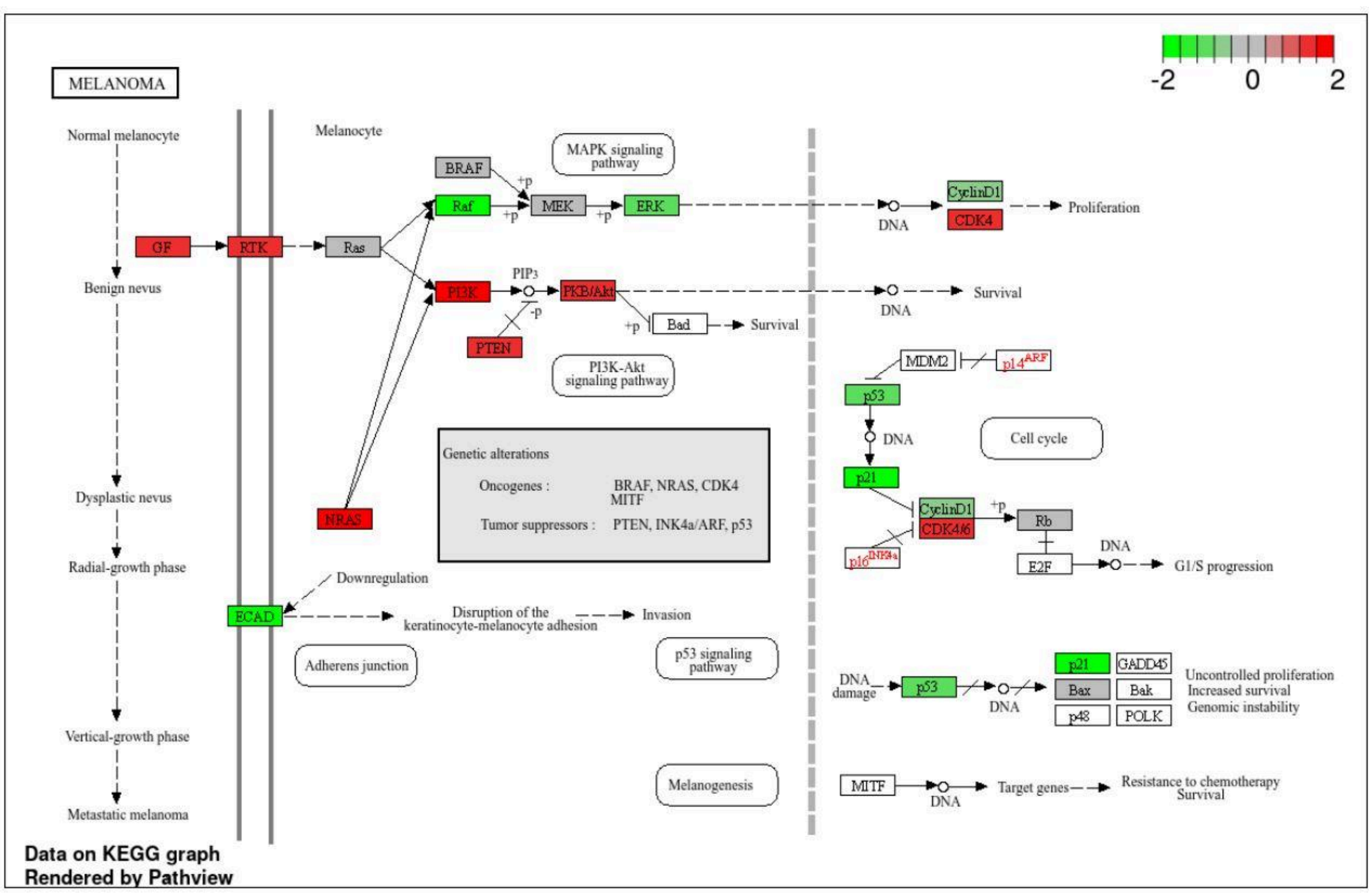

**Fig 4. Expression profiles of cell-cycle related genes visualized on a melanoma KEGG signaling pathway diagram using Pathview.** Red and green shading indicate upregulated and downregulated genes, respectively.

## Mathematical Model

In this section, we present our reduced "within-host" mathematical model of tumor and immune system dynamics. Our model is presented as a system of ODEs that describes the interactions between melanoma tumor cells (M), M1 macrophages (M1), M2 macrophages (M2), cytotoxic T cells ($T_C$), and regulatory T cells ($T_R$). The M1 and M2 macrophage populations are critical to the innate immune response, whereas $T_C$, and $T_R$ are central to the adaptive immune response. The interactions between the cell populations included in this model are illustrated by arrows in **Fig 5**. Details about the biological basis for each arrow, or pathway, are given in the next section. The descriptions of model pathways are summarized in **Table 5**. These pathways represent "net" effects in the system. Features that are not explicitly modeled, such as other pathways, or dendritic cells, and B cells, likely contribute to these "net" effects.

## Description of model dynamics

In this section, we provide details of the mathematical model represented by the diagram in **Fig 5** and by equations (1)-(5). We will use the term "arrow" to refer to curves with arrowheads at the end, (indicating a boosting effect) as well as for curves with perpendicular line segments at the end (indicating a damping effect). Red arrows signify anti-tumor effects; blue arrows signify pro-tumor effects. As can be seen in the equations, many of the rate effects we describe in this section use terms with a Michaelis-Menten style. This is to ensure that the terms we use do not result in unbounded rates. Each cell in the model has a loss rate, which we assume is proportional to the cells present. This includes the loss of M1 to become M2, and vice versa.

**Melanoma (M):** The model is centered around the population of melanoma cells, labeled M in **Fig 5**. The melanoma cells proliferate, which is indicated by the black circular arrow above the M compartment. We assume that the tumor cells proliferate logistically with rate constant $r_M$, as experimental data have shown that the tumor growth is eventually slowed down in the case of lack of nutrients.[71] The presence of M2 macrophages promotes tumor growth (represented by pathway **f** multiplying logistic growth in Eq. 1) via low expression of IL-12, and high expression of IL-10, IL-1 decoy receptor, and IL-1 receptor antagonist.[72] The natural loss of M is represented by the black downward arrow pointing away from the cell population. Cytotoxic T cells kill tumor cells (represented by pathway **k**) after recognizing antigens on the tumor cells, and cause apoptosis via release of granzymes, perforin, cathepsin C, and granulysin into the tumor cell.[73] Cytotoxic T cells can also induce apoptosis of tumor cells by expressing Fas ligand which binds with Fas on tumor cells.[74] However, M2 cells, via increased expression of PD-L1 and IL-10, reduce the effectiveness of $T_C$ cells killing melanoma cells (represented by the **g** pathway) by creating an immunosuppressive environment in the tumor.[75] . Additionally, $T_R$ cells suppress the effect of $T_C$ cells,[9] (represented by the **l** pathway), which is accomplished by secretion of cAMP, adenosine, IL-10, IL-35, and TGF-β.[77] The cancer cells (M) also contribute to $T_c$ cell dysfunction via PD-L1 and PD-1 interactions (shown as pathway j).[78]

**M1 Macrophages (M1):** In the microenvironment of a solid tumor, macrophages play a crucial role, which could result in either tumor progression or tumor suppression,[72,79] both of which can be affected by therapies.[79,80] We assume that there is a constant source of M1 macrophages, and that this is more significant than proliferation, similar to what has been assumed in other models.[25] In response to signaling pathways, M1 macrophages can polarize to M2 macrophages (represented by pathway **c**) and vice versa (represented by pathway **b**), changing their function. [80] Boddaert et al. showed that $T_C$ cells are able to repolarize M2 macrophages to M1 (represented by pathway **a**)[82]. The presence of tumor cells increases the rate of polarization of M1 to M2 (represented by pathway **e**),[83] specifically by TGF-β.[84]

**M2 Macrophages (M2):** M2 macrophages are recruited to the tumor site, where they release tumor growth-promoting agents and other cytokines (represented by pathway **f**), including IL-10, ADM, IFN-γ, angiotensin, COX-2, and IL-1β. [85,86] We assume that the proliferation term is negligible in comparison to the constant source (represented by the black arrow pointing toward M2, with rate constant $s_2$), as in other models.[81] The presence of tumor cells increases the rate of polarization to M2, reducing the number of M1 cells; this process is represented by multiplying **c** by **e**. Pathway **b** represents the repolarization of M2 macrophages to M1, and pathway **a** represents the effect of $T_C$ cells in converting M2 macrophages to M1.

**Cytotoxic T lymphocytes ($T_C$):** Another critical immune cell type we include in our mathematical model is the population of activated cytotoxic T cells ($T_C$). CD8+ T cells in the TME have been shown to positively correlate with a good prognosis for success with cancer immunotherapy.[87] Activated cytotoxic CD8+ T cells ($T_c$ induce apoptosis of cancer cells by releasing cytotoxic granules,[88] which is represented by arrow **k**.[89] Further, CD8+ T cells can release IFN-γ, which can induce apoptosis of cancer cells along with granzyme B and perforin.[90] Activated $T_C$ cells proliferate, which is represented by the circular arrow on top of the $T_C$ population in **Fig 5**. We assume this proliferation is logistic (Eq. 4). M1 macrophages present antigens from phagocytized tumor cells to $T_C$ cells, resulting in $T_C$ activation, and subsequent proliferation, parts of the adaptive immune response.[83] Since the antigen presentation process and its downstream proinflammatory effects are dependent on the both M1 macrophages and melanoma cells, we have represented this interaction using pathways **d** and **i**, respectively. The loss of $T_C$ cells, represented by the black arrow starting from $T_C$ cells and pointing away, is assumed to capture both death of $T_C$ cells and exhaustion of their functions.

**Regulatory T cells ($T_R$):** Regulatory T cells are another essential component of our model. Their proliferation is represented by the circular arrow on top of the $T_R$ population, and we assume the growth to be logistic (Eq. 5). $T_R$ cells are known to suppress CD8+ T cell function through TGF-β;[76,91] this is represented by pathway **l**. M2 macrophages express chemokines including CCL20, CCL22, and others,[92] which have been shown to support the development of Tregs in a tumor microenvironment;[93,94] this is represented by **h** in **Fig 5**.

## Parameters

Some parameter values were found in the literature, based on in vitro or in vivo experiments. Others were derived from other mathematical models or estimated from clinical experience. We calculated initial values for melanoma cells and immune cells based on the experimental study of Erdag et al.[95] The list of parameters, descriptions, estimated values, and sources is provided in **Table 6**.

In addition to representing the specific effects included in our model descriptions, we note again that parameters capture "net" effects of biological processes not explicitly modeled. These net effects could incorporate effects from other cells not included explicitly in our model, such as B cells, natural killer cells, dendritic cells, or other immune cells. Interactions that have not been explicitly included between modeled populations, but are known to occur, could also contribute to net effects. This means that when the model is fit to in vivo data, the estimated values of model parameters will incorporate these net effects. These parameter estimates would be different if we expanded the model and had some of these other effects explicitly incorporated into additional model terms.

### Initial values

To calculate the initial values for our model, we used data and formulas from Erdage et. al.[95] They selected 183 metastatic melanoma tissue samples from 147 patients. We employed their methodology to calculate the number of immune cells and melanoma cells. They used a simple formula for cell density based on the number of cells observed in a thin slice of tissue:

$$\frac{\text{Cells}}{\text{mm}^3} = \frac{\text{number of cells}}{\text{mm}^2} * \frac{1}{\text{diameter (in mm)}}$$

. Since Erdage et. al assumed a diameter of 10 microns, and 10 microns or 0.01 millimeters, for immune cells, we have:

$$\frac{\text{Immune cells}}{\text{mm}^3} = \frac{\text{number of cells}}{\text{mm}^2} * \frac{1}{0.01 \text{ mm}} = \frac{\text{number of cells}}{\text{mm}^2} * \frac{100}{\text{mm}}.$$

For melanoma cells, they used the same formula and assumed a diameter of 20 microns. Immunotype B is close to the steady-state phase of the cancer, so we calculated the values of immune cells and melanoma cells in the steady state phase, using values from Table 2. This gave the following estimates for initial values:

12580 $CD8^+$ T cells/$\text{mm}^3$,

2540 Tregs/$\text{mm}^3$,

4740 M2 macrophages/$\text{mm}^3$. Assuming the ratio of M2/M1 is 1.75, we get 2709 M1 macrophages/$\text{mm}^3$.

We calculate the density of melanoma cells by the following formula: 12580*2 = 25160 cells per $\text{mm}^3$.

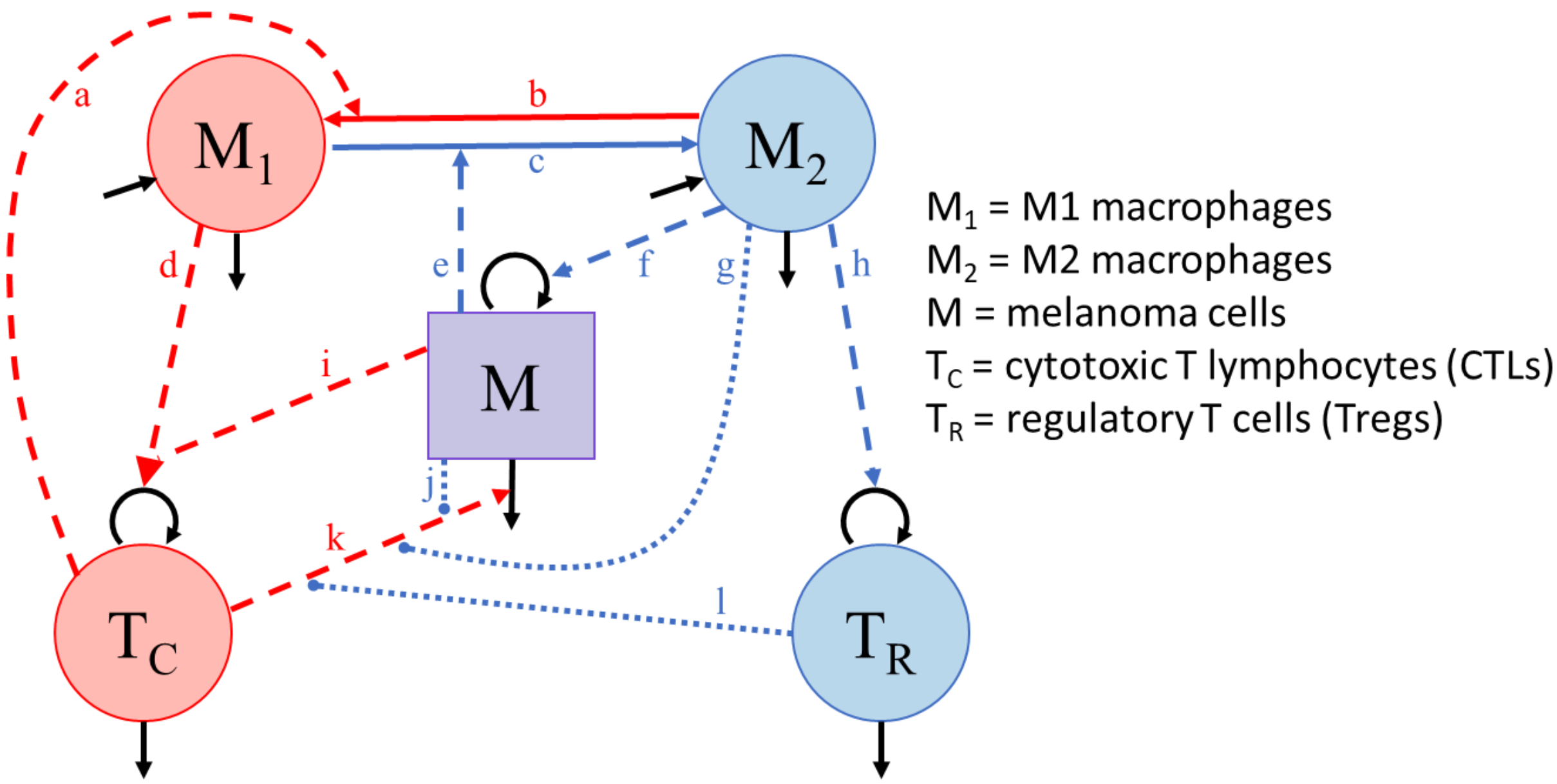

**Fig 5. Diagram of population interactions.** $M_1$ represents M1 macrophages, $M_2$ represents M2 macrophages, M represents melanoma cells, $T_C$ represents cytotoxic T lymphocytes (CTLs), and $T_R$ represents regulatory T cells (Tregs). The solid curves represent an increase (arrows pointing toward a population) or decrease (arrows pointing away from a population) in population size. The dashed arrows represent interactions that boost the mechanisms they are pointing to. The dotted curves, with filled circular end points, denote interactions that inhibit the mechanisms

they are pointing to. Red arrows and labels represent anti-tumor immunologically hot components, while blue represents pro-tumor immunologically cold components. The purple of the melanoma represents the result of both of these types of effects.

| Pathway | Description | References |
|---|---|---|
| a | $T_C$ boosts repolarization of M2 macrophages to M1 | [82] |
| b | M2 macrophages repolarize to M1 | [80] |
| c | M1 macrophages repolarize to M2 | [80,96], [96] |
| d | M1 macrophages boost $T_C$ proliferation | [83] |
| e | M cells boost M1 → M2 repolarization | [83] |
| f | M2 macrophages boost M proliferation | [93], [97],[98], [99] |
| g | M2 macrophages decrease $T_C$ efficacy for killing M cells | [82], [100], [101] |
| h | M2 macrophages boost $T_R$ proliferation | [93] |
| i | Antigen shed by M cells boosts $T_C$ proliferation | [102] |
| j | M cells decrease $T_C$ efficacy for killing M cells | [78] |
| k | $T_C$ cells kill melanoma cells | [103], [99] |
| l | $T_R$ decreases $T_C$ efficacy in killing M cells | [76] |

**Table 5. Summary of the 12 cellular pathways included in the model diagram.** Red color indicates the anti-tumor pathways, and blue indicates pro-tumor pathways.

The dynamics of the cell populations in the model are given by Equations (1) - (5). The parameter descriptions and values are shown in **Table 6.**

$$
\begin{aligned}
\frac{dM}{dt} &= r_M\left(1-\frac{M}{K_M}\right)M\overbrace{\left(1+\frac{\alpha_{2M}M_2}{\beta_{2M}+M_2}\right)}^{f} \\
&\quad -\delta_M M\left(1+\overbrace{\left(\frac{\alpha_{CM}T_C}{\beta_{CM}+T_C}\right)}^{k}\overbrace{\left(1-\frac{\alpha_{2CM}M_2}{\beta_{2CM}+M2}\right)}^{g}\overbrace{\left(1-\frac{\alpha_{MCM}M}{\beta_{MCM}+M}\right)}^{j}\overbrace{\left(1-\frac{\alpha_{RCM}T_R}{\beta_{RCM}+T_R}\right)}^{l}\right),
\end{aligned} \tag{1}
$$

$$
\frac{dM_1}{dt} = s_1+\overbrace{(\gamma_2 M_2)}^{b}\overbrace{\left(1+\frac{\alpha_{C21}T_C}{\beta_{C21}+T_C}\right)}^{a}-\overbrace{(\gamma_1 M_1)}^{c}\overbrace{\left(1+\frac{\alpha_{M21}M}{\beta_{M21}+M}\right)}^{e}-\delta_1 M_1, \tag{2}
$$

$$
\frac{dM_2}{dt} = s_2+\overbrace{(\gamma_1 M_1)}^{c}\overbrace{\left(1+\frac{\alpha_{M12}M}{\beta_{M12}+M}\right)}^{e}-\overbrace{(\gamma_2 M_2)}^{b}\overbrace{\left(1+\frac{\alpha_{C21}T_C}{\beta_{C21}+T_C}\right)}^{a}-\delta_2 M_2, \tag{3}
$$

$$
\frac{dT_C}{dt} = r_C\left(1-\frac{T_C}{K_C}\right)T_C\left(1+\overbrace{\left(\frac{\alpha_{1C}M_1}{\beta_{1C}+M_1}\right)}^{d}\overbrace{\left(\frac{\alpha_{M1C}M}{\beta_{M1C}+M}\right)}^{i}\right)-\delta_C T_C, \tag{4}
$$

$$
\frac{dT_R}{dt} = r_R\left(1-\frac{T_R}{K_R}\right)T_R\overbrace{\left(1+\frac{\alpha_{2R}M_2}{\beta_{2R}+M_2}\right)}^{h}-\delta_R T_R. \tag{5}
$$

| # | Name | Pathway | Description | Value | Units | References |
|---|---|---|---|---|---|---|
| 1 | $\alpha_{C21}$ | **a** | Maximum boost by $T_C$ on M2 to M1 conversion | 5 | - | Estimated |
| 2 | $\beta_{C21}$ | **a** | Threshold for half of maximum boost by $T_C$ on M2 to M1 conversion | 1500 | cells/$\mu$L | Estimated |
| 3 | $\gamma_2$ | **b** | M2 to M1 transition rate constant | 0.04 | 1/day | [[104]] |
| 4 | $\gamma_1$ | **c** | M1 to M2 conversion rate constant | 0.2 | 1/day | [[105]] |
| 5 | $\alpha_{1C}$ | **d** | Maximum boost on $T_C$ proliferation by M1 antigen presentation | 2.1 | - | Estimated |
| 6 | $\beta_{1C}$ | **d** | Threshold for half of maximum boost of M1 antigen presentation on $T_C$ proliferation | 5000 | cells/$\mu$L | Estimated |
| 7 | $\alpha_{M12}$ | **e** | Maximum boost by M on M1 to M2 conversion | 0.4 | - | Estimated |
| 8 | $\beta_{M12}$ | **e** | Threshold for half of maximum boost by M on M1 to M2 conversion | 1000 | cells/$\mu$L | Estimated |
| 9 | $\alpha_{2M}$ | **f** | Maximum boost by M2 on M proliferation | 0.3 | - | Estimated |

| 10 | $\beta_{2M}$ | **f** | Threshold for half of maximum boost by M2 on M proliferation | 1500 | cells/$\mu$L | Estimated |
|---|---|---|---|---|---|---|
| 11 | $\alpha_{2CM}$ | **g** | Maximum inhibition by M2 on $T_C$ efficacy for killing M | 0.8 | - | Estimated |
| 12 | $\beta_{2CM}$ | **g** | Threshold for half of maximum inhibition by M2 on $T_C$ efficacy for killing of M | 1000 | cells/$\mu$L | Estimated |
| 13 | $\alpha_{2R}$ | **h** | Maximum boost by M2 on $T_R$ proliferation | 0.5 | - | Estimated |
| 14 | $\beta_{2R}$ | **h** | Threshold for half of maximum boost by M2 on $T_R$ proliferation | 1000 | cells/$\mu$L | Estimated |
| 15 | $\alpha_{M1C}$ | **i** | Maximum boost by M on M1 boost of $T_C$ proliferation | 0.075 | - | Estimated |
| 16 | $\beta_{M1C}$ | **i** | Threshold for half of maximum boost by M on M1 boost of $T_C$ proliferation | 1000 | cells/$\mu$L | Estimated |
| 17 | $\alpha_{MCM}$ | **j** | Maximum inhibition by M on $T_C$ efficacy for killing M | 0.2 | - | Estimated |
| 18 | $\beta_{MCM}$ | **j** | Threshold for half of maximum inhibition by M on $T_C$ efficacy for killing M | 1500 | cells/$\mu$L | Estimated |
| 19 | $\alpha_{CM}$ | **k** | Maximum boost by $T_C$ on M loss rate | 5 | - | [106] |
| 20 | $\beta_{CM}$ | **k** | Threshold for half of maximum boost by $T_C$ on M loss rate | 1000 | cells/$\mu$L | Estimated |
| 21 | $\alpha_{RCM}$ | **l** | Maximum inhibition by $T_R$ on $T_C$ efficacy for killing M | 0.25 | - | [106] |
| 22 | $\beta_{RCM}$ | **l** | Threshold for half of maximum inhibition by$T_R$ on $T_C$ efficacy for killing M | 500 | cells/$\mu$L | Estimated |
| 23 | $r_M$ | | Tumor growth rate constant | 0.431 | 1/day | [107],[71] |
| 24 | $K_M$ | | Carrying capacity of tumor | 1.00E+09 | cells/$\mu$L | [105] |
| 25 | $\delta_M$ | | Tumor natural death rate constant | 0.17 | 1/day | Estimated |
| 26 | $s_1$ | | Constant source rate for M1 | 270 | cells/ ($\mu$L*days) | Estimated |
| 27 | $\delta_1$ | | Death rate constant for M1 | 0.2 | 1/day | [104] |
| 28 | $s_2$ | | Constant source rate for M2 | 948 | cells/ ($\mu$L*days) | Estimated |
| 29 | $\delta_2$ | | Death rate constant for M2 | 0.08 | 1/day | [108] |
| 30 | $r_C$ | | Proliferation rate constant for $T_C$ | 0.2 | 1/day | [109] |

| 31 | $K_C$ | | Carrying capacity for $T_C$ | 35000 | cells/$\mu$L | Estimated |
|---|---|---|---|---|---|---|
| 32 | $\delta_C$ | | Loss/inactivation rate constant for $T_C$ | 0.2 | 1/day | [109] |
| 33 | $r_R$ | | Proliferation rate constant for $T_R$ | 0.02 | 1/day | Estimated |
| 34 | $K_R$ | | Carrying capacity for $T_R$ | 25000 | cells/$\mu$L | Estimated |
| 35 | $\delta_r$ | | Loss/inactivation rate constant for $T_R$ | 0.014 | cells/$\mu$L | [110] |
| 36 | $M_0$ | | Initial value of melanoma cells | 25160 | cells/$\mu$L | [95] |
| 37 | $M1_0$ | | Initial value of M1 macrophages | 2709 | cells/$\mu$L | [95] |
| 38 | $M2_0$ | | Initial value of M2 macrophages | 4740 | cells/$\mu$L | [95] |
| 39 | $T_{C0}$ | | Initial value of $T_C$ | 12580 | cells/$\mu$L | [95] |
| 40 | $T_{R0}$ | | Initial value of $T_R$ | 2540 | cells/$\mu$L | [95] |

**Table 6. Parameter descriptions with nominal values ("Value") used in the modeling.** The "#" column indicated the parameter identification number.

**Numerical simulation**

We numerically solved the system using MATLAB 2024 to understand the dynamics of the cell populations in the model. For these simulations, we used parameter values that are listed in the "Value" column in **Table 6**. The simulation results are shown in **Fig 6**. M1 macrophages are continuously recruited, but they are also converted to M2 macrophages. Due to the presence of inflammation, some M1 macrophages do not convert to M2. The conversion of M1 macrophages to M2 macrophages boosts the levels of M2. The melanoma cells grow quickly to their resource-constrained limit, which is imposed by a logistic term that limits their growth. In this later phase, $T_C$ cells are not as effective and they enter an exhausted state. The plot of $T_R$ agrees with experimental studies that show that changes associated with tumor growth (e.g., altered nutrient composition and oxygen availability, cytokines released by tumor cells, stroma, and immune cells) also favor Treg infiltration and effector T cell exhaustion.[111]

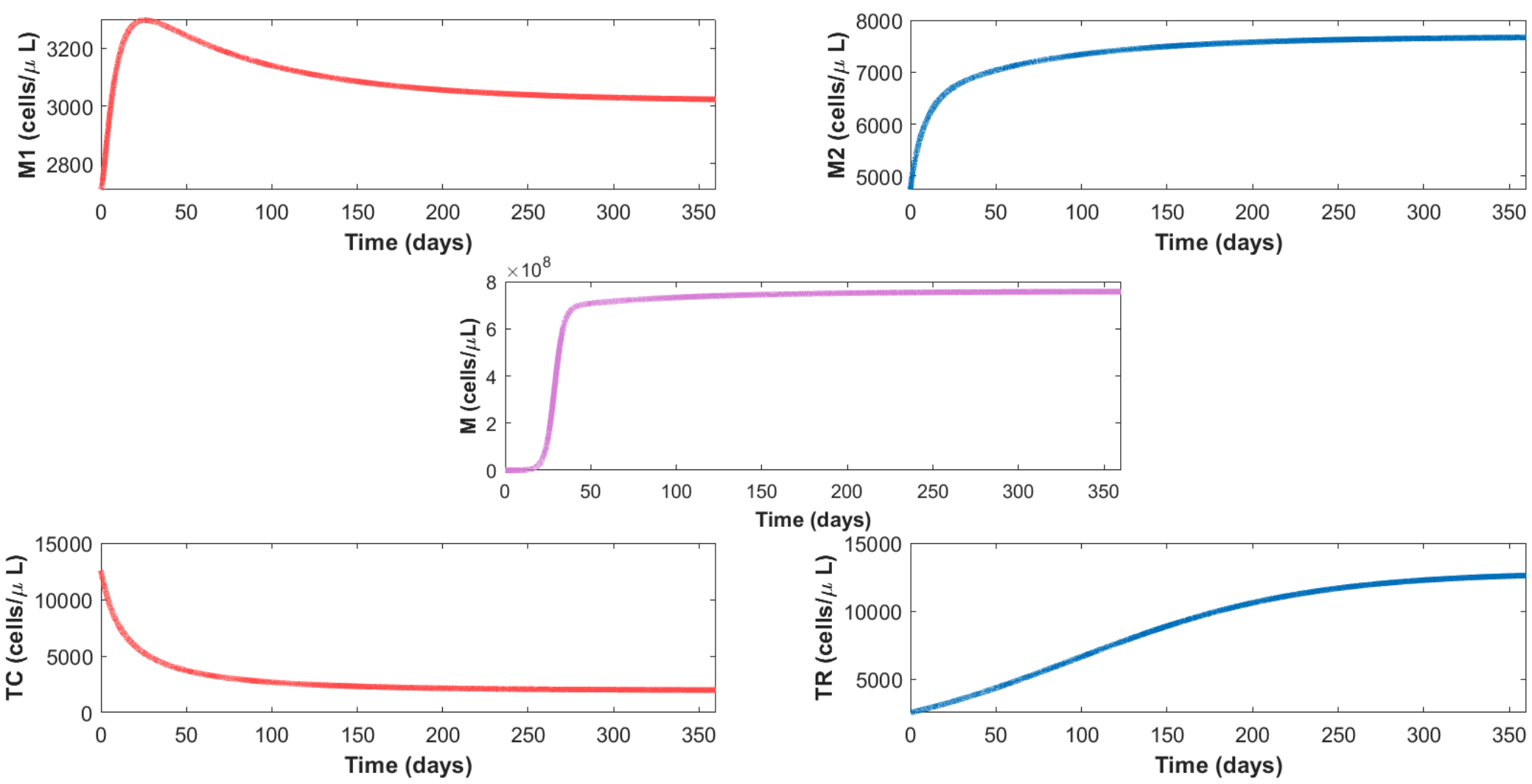

**Fig 6. Melanoma disease burden and immune cell levels in the absence of treatment.** These simulated populations use the parameter values and initial conditions listed in Table 6. Melanoma burden is measured using the longest diameter (mm) and all immune cells are measured in cells/µL or cells/mm$^3$ , which represent equivalent volumes. The horizontal axes in all of these plots are in units of time (days).

In our sensitivity analysis results, shown in **Fig 7**, we observed that $K_M$, $r_M$, $\delta_M$, $\delta_C$, $r_C$, and $\alpha_{2CM}$ are the parameters that have the most influence on the tumor burden at day 180. Since parameters describing interactions with $T_C$ have been shown to be crucial, there has been a significant focus on immunotherapy in cancer by restoration of or boosting the CD8+ T cell function in melanoma therapies.[99,103] While these CD8+ T cell-based therapies have shown significant benefit, other new cancer therapies involving CD8+ T cell immunotherapy have also shown great potential, including neoantigen vaccination, chimeric antigen receptor T cell (CAR-T) and T-cell receptor T (TCR-T) therapy, and immune checkpoint blockers.[92] The FDA has approved the use of autologous cell therapy using tumor infiltrating lymphocytes (TILs) in the treatment of melanoma that is metastatic or unresectable.[112]

Other immunotherapeutic strategies have been proven efficacious as well, such as the use of IL-2 therapy to help stimulate and activate T cells. The bioengineering of TCR expression for specific tumor antigen or epitopes on CD8+ T Cells of cancer patients is another therapeutic strategy.[113] It can then be postulated that the sensitivity analysis results for parameters associated with activation, proliferation, and T cell destruction in our model are the most important, and may support experimental investigations and clinical conclusions.

The parameter $\alpha_{2CM}$, in pathway g, was also identified as highly influential. Our simulations are consistent with clinical experiments that show M2 macrophages inhibit CD8+ T cells from reaching tumor cells and limit the efficacy of anti-PD-1 treatment.[100] Therefore, interactive pathways between CD8+ T cells and M2 macrophages may be potential targets for immunotherapy.

The top most influential parameters, $K_M$, $r_M$, and $\delta_M$ show the outcome is highly sensitive to growth or death of melanoma cells. This result agrees with clinical and experimental experience with chemotherapies and targeted therapies that have already been used to treat melanoma by killing melanoma cells or stopping proliferation, which reflects the importance of these influential parameters.[114]

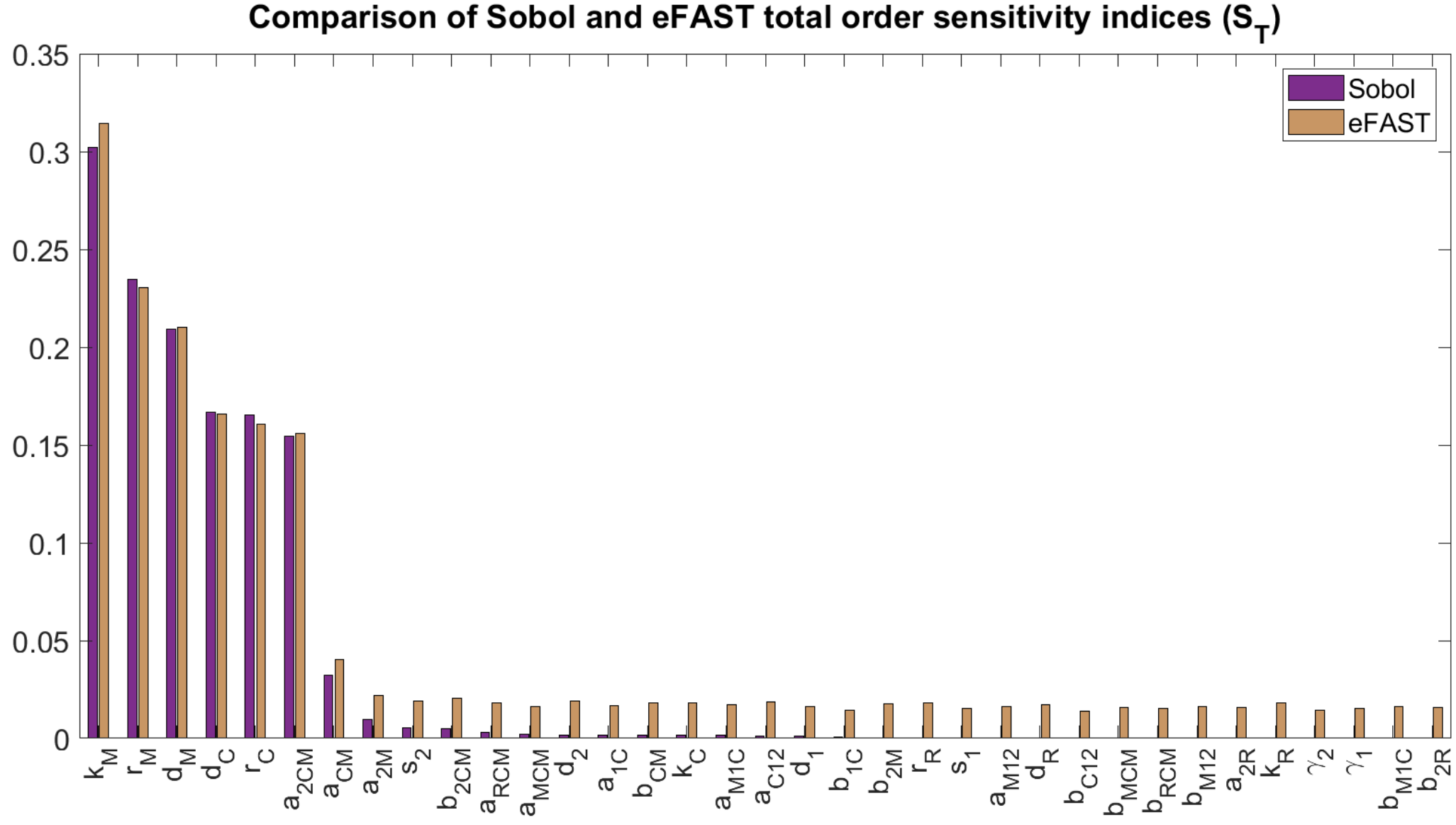

**Fig 7. Sobol and eFAST total-order sensitivity indices for model parameters.** Parameters are ranked based on Sobol total-order index value, $S_T$, which illustrates the sensitivity of the QOI to both an individual parameter's effects and the effects from its interactions with other parameters. The top most-influential parameters are $K_M$, $r_M$, $\delta_M$, $\delta_C$, $r_C$, and $\alpha_{2CM}$. Results were obtained by using a base sample size of 100,000.

| Top-Ranked Parameters | Rank (both Sobol and eFAST) | Sobol $S_T$ | eFAST $S_T$ |
|---|---|---|---|
| $K_M$ | 1 | 0.3022 | 0.3145 |
| $r_M$ | 2 | 0.2349 | 0.2302 |
| $\delta_M$ | 3 | 0.2094 | 0.2103 |
| $\delta_C$ | 4 | 0.1665 | 0.1656 |
| $r_C$ | 5 | 0.1650 | 0.1608 |
| $\alpha_{2CM}$ | 6 | 0.1543 | 0.1559 |

**Table 7. Top six total sensitivity index values from Sobol and eFAST methods.** The ranking is the same for both methods for these top six parameters. These were computed with the same 100,000 base samples (675,000 total samples) used for **Fig 7**.

**Kolmogorov-Smirnov Test**

To ensure that the six parameters that were determined to be most influential (according to the Sobol total-order index) adequately capture the variability of the QOI, we used the two-sample Kolmogorov-Smirnov (KS) test. The first sample is the distribution of the QOI values for which we allow all parameters to vary. The second sample is the distribution of the QOI values for which we allowed the six most-influential parameters to vary and fix the 29 least-influential parameters at their nominal values. The KS p-value for the comparison of the distributions was 3.9e-55. Using a significance level of 0.05, these two samples very likely come from different distributions. For larger samples, the KS test becomes better at detecting even minor differences between distributions. Performing a sensitivity analysis on a model with 35 parameters and 175,000 base samples yields 6,475,000 model evaluations, each resulting in a final QOI value. Varying only the top 6 parameters would yield 1,400,000 model evaluations and QOI values. The KS test statistic is influenced by how large each sample is, and both samples of size 6,475,000 and 1,400,000 are extraordinarily large. To visualize the difference between the two distributions of model outputs, **Fig 8** shows the histogram of QOI values calculated by letting all parameters vary (red) is visually similar to the histogram of QOI values calculated by only letting the top six most-influential parameters vary (black). Despite a low p-value from the KS test, **Fig 8** gives us confidence that the six most-influential parameters do capture most of the variability in the QOI. Also, we plan to restrict the number of pathways targeted with combination therapy to six or fewer. Thus we continue to restrict our attention to the top six parameters.

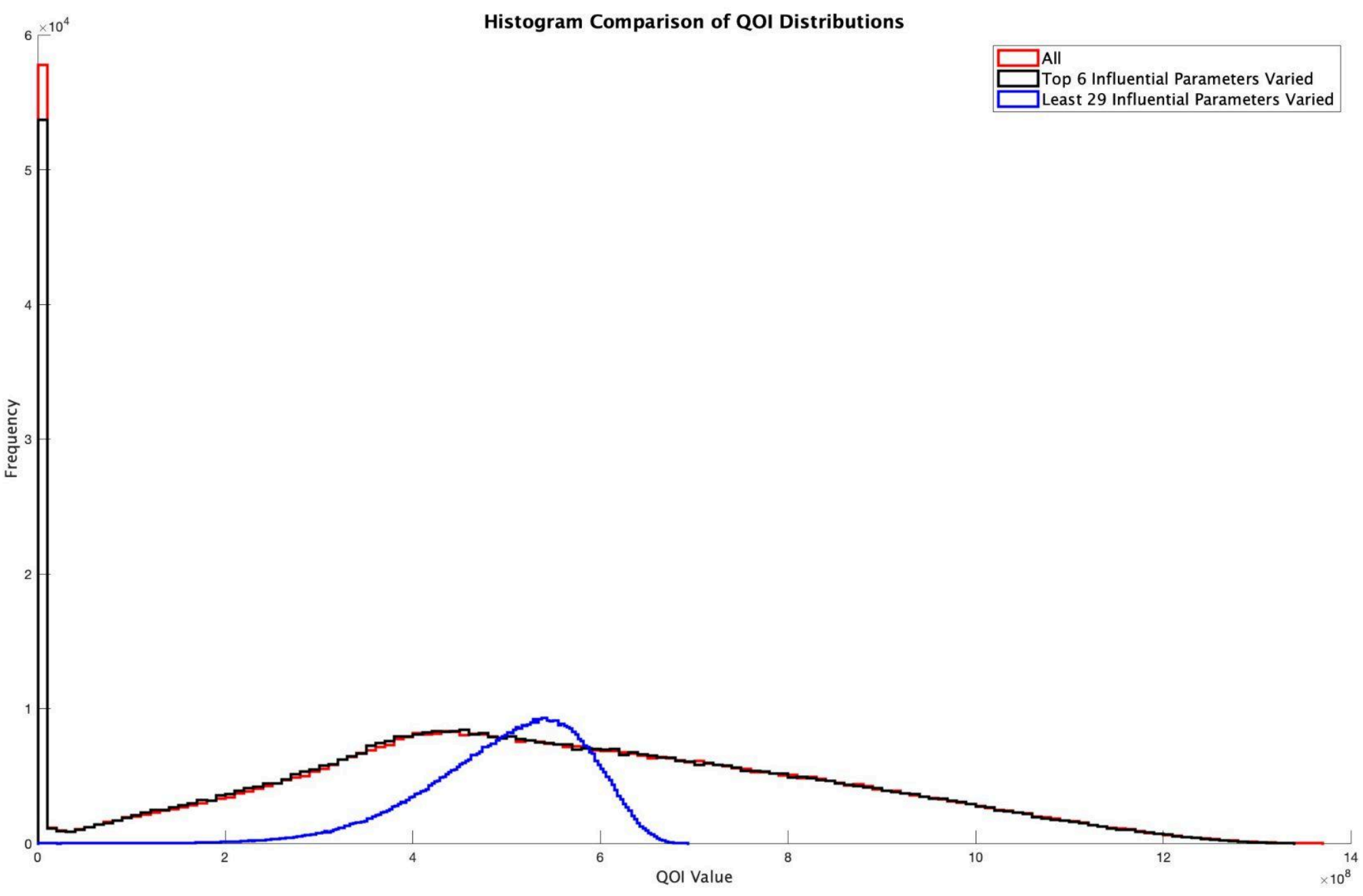

**Fig 8. Comparison of histograms of QOI values resulting from different sets of parameters.** Similar to the sensitivity analysis procedure, we alloted each parameter a uniform distribution, with a lower bound of half its nominal value, and an upper bound of 1.5 times its nominal value. To plot the red curve, we sampled from each parameter's distribution 300,000 times using Sobol sequencing, resulting in 300,000 parameter sets. We solved the model with each of these parameter sets, and plotted the QoI in red. Since all parameters were allowed to vary, we name this the "full-model" behavior. To visualize how much variability in the QOI is captured by the top six most-influential parameters, we used the same sampling scheme for parameter sets with only the top six influential parameters varying, and reevaluated the model. The black curve was produced by plotting all QOI values obtained by letting the top six parameters vary and keeping the 29 least-influential parameters fixed at their nominal values. The black curve captures much of the full-model behavior. The blue curve was obtained keeping the six top-influential parameters fixed and letting the 29 least-influential parameters vary. The blue curve notably does not capture full-model behavior. This figure was produced using MATLAB, version R2024b.

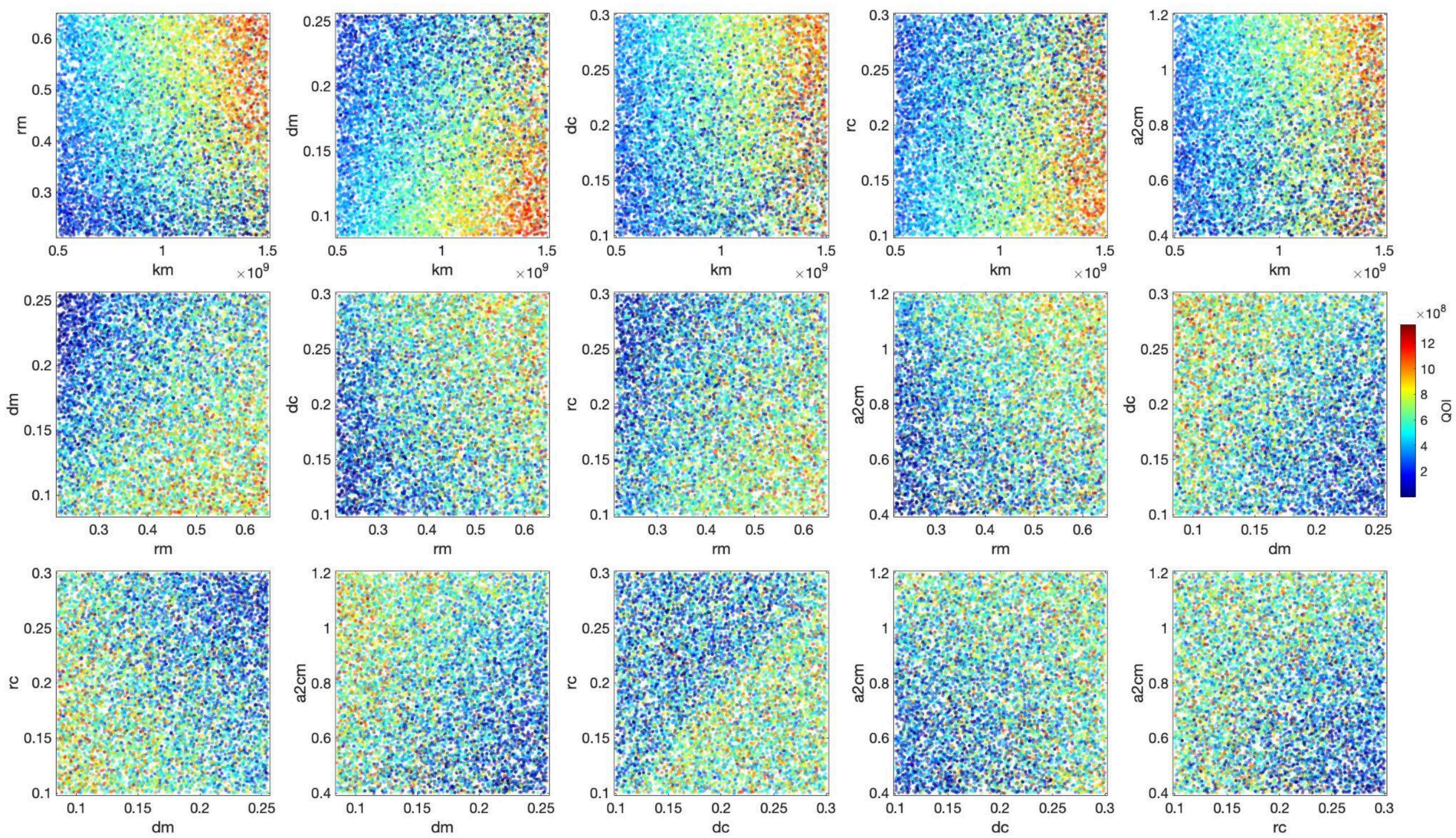

**Fig 9. Pairwise scatterplots of the six most-influential parameters and QOI values.** Each scatterplot shows how the QOI values are arranged with regard to values of the six most-influential parameters. These plots were produced using a subset of 10,000 evaluations randomly selected from the 6,475,000 model evaluations in the original sensitivity analysis. The points are colored by magnitude of QOI values, with red representing highest values and blue representing lowest values of the QOI. Each scatterplot in the top row shows clear separation between blue and red; the other scatterplots have more mixing between the colors, but still show some separation of colors. This agrees with the rank ordering of the top-influential parameters ( $K_M$, $r_M$, $\delta_M$, $\delta_C$, $r_C$, $\alpha_{2CM}$), and provides more-detailed insight into how these parameters drive the QOI values. This figure was produced using MATLAB, version R2024b.

**Identifiability Analysis**

To determine if any subset of the top six most-influential parameters could be uniquely estimated from data corresponding to the cell populations M, M1, M2, $T_C$ and $T_R$ , we performed structural identifiability analysis. When considered together, the top 6 parameters are only locally identifiable. In other words, if we fit our model to data, these parameters cannot be uniquely estimated, and could take any of a finite number of values.

We then tested just $K_M$, $r_M$, $\delta_C$, $r_C$, $\alpha_{2CM}$ for structural identifiability, keeping $\delta_M$ and the other 29 parameters fixed at their nominal values. The results indicated that these 5 influential parameters are globally identifiable, and can be uniquely estimated with adequate data. The algorithm computes derivatives of the original system to obtain new equations, and tests to see if they are independent. The tableau in **Fig 10A** represents these new equations in the rows, with each column corresponding to a parameter. A black entry indicates there is a non-zero coefficient of the parameter in that equation, and a white entry indicates the parameter does not appear in that equation. The corresponding reduced tableau (**Fig 10B**) shows the independent equations. Since there are five independent equations, all five parameters are globally identifiable.

A B

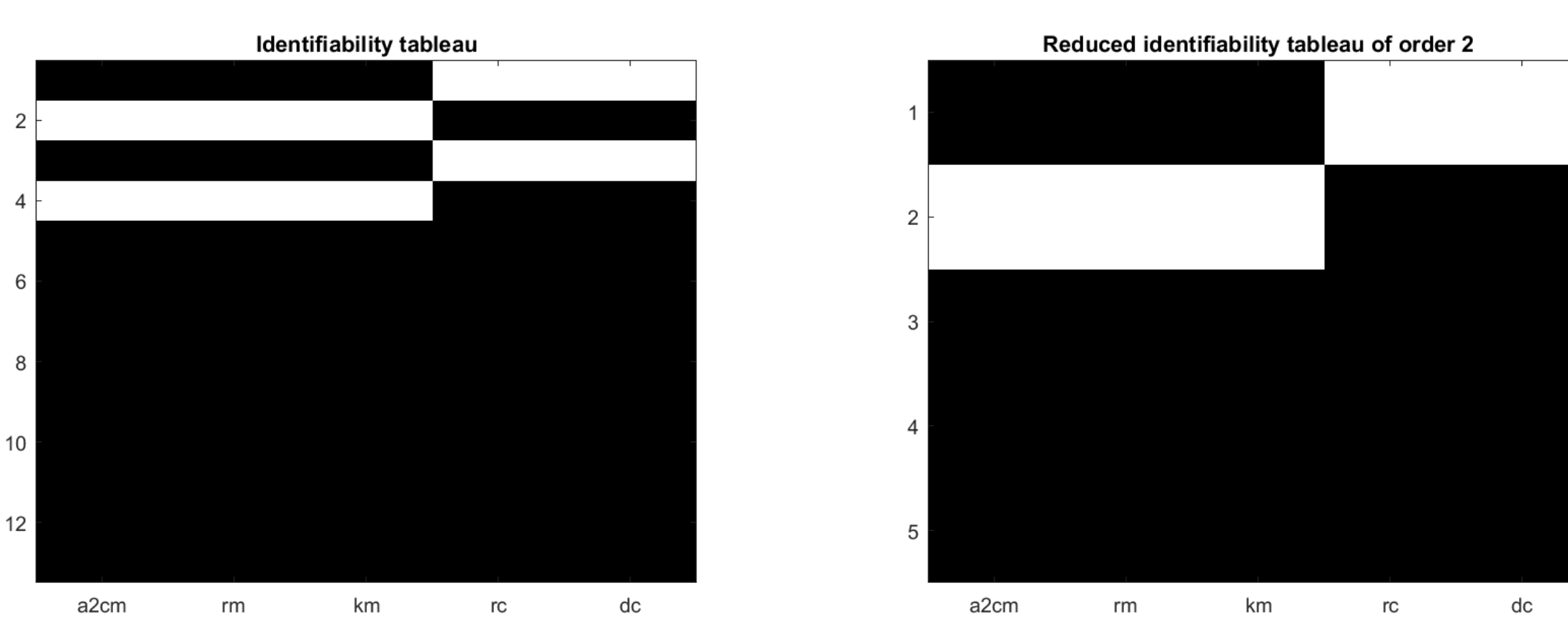

**Fig 10. Model identifiability tableaus generated with GenSSI.** (A) Initial full tableau. (B) Reduced tableau. Both tableaus considered five of the six top most-influential parameters, omitting $\delta_M$. Each row represents an equation, and each column is a parameter. Black indicates that the equation has a non-zero coefficient for the given parameter. GenSSI generates equations by using Lie derivatives of the original model system, and checks to see if there are enough independent equations to uniquely identify all the specified parameters. All of the five included parameters ($K_M$, $r_M$, $\delta_C$, $r_C$, $\alpha_{2CM}$) are globally identifiable, which means they can be uniquely estimated given adequate data.

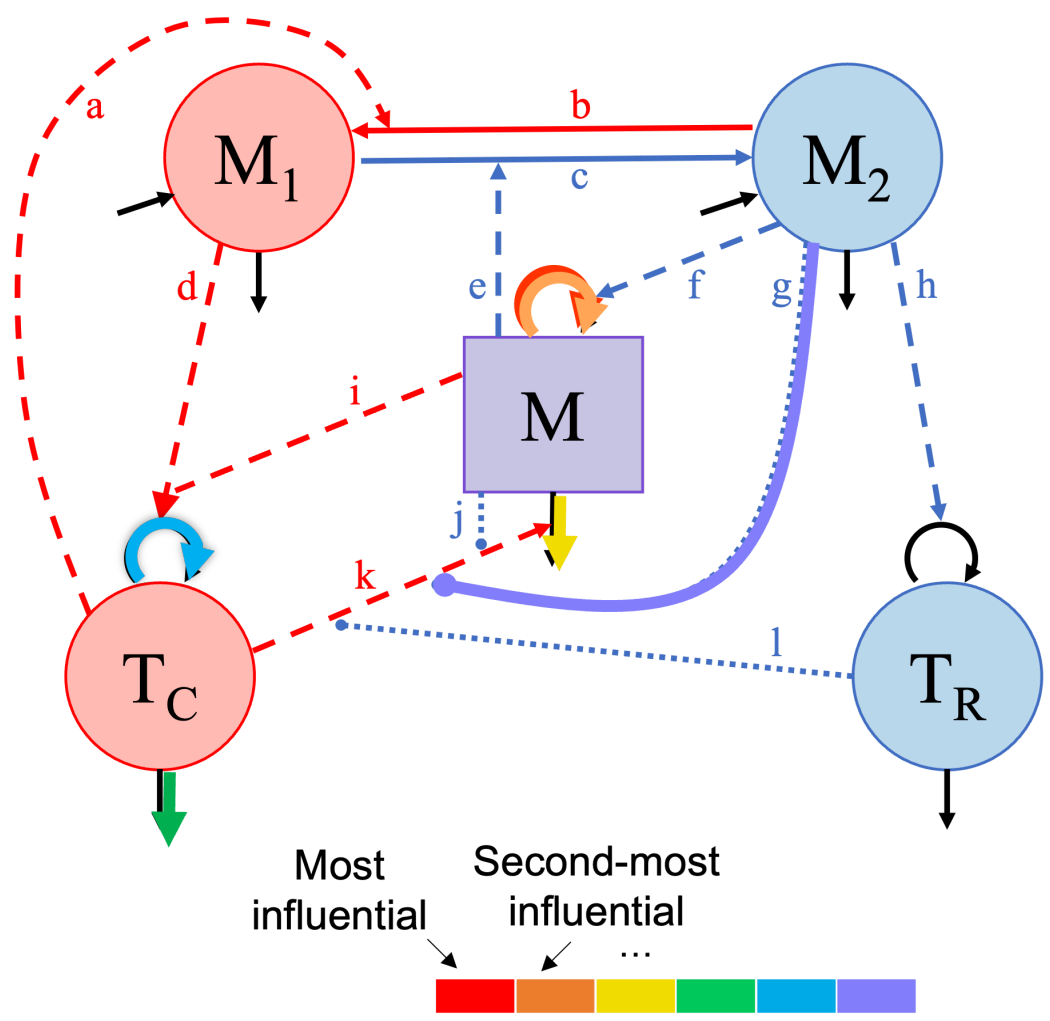

**Fig 11. Heat map showing sensitivity analysis results for the melanoma model.** The most-influential parameter is included in a pathway that is shown in red, the second-most influential parameter is in a pathway shown in orange (same pathway as the red), the third-most influential parameter is in the pathway shown in yellow, etc.

## Discussion

In this work, we developed a new mechanistic mathematical model to help understand the tumor-immune microenvironment dynamics in patients with rare melanomas. Our initial mathematical model was based on literature information and included melanoma cells and 15 different immune cell types. Because of challenges associated with a model of this size, we used data from canine melanoma patients to reduce the model while retaining key tumor-immune dynamics.

Spontaneously-arising canine melanoma provides a valuable comparative oncology model for rare human melanomas. We collected primary melanoma tumor samples matched with metastatic lymph node tumor samples from eight canine melanoma patients. We also collected healthy canine tissue samples from four healthy canines to use for comparison.

We used the information obtained from the tissue samples to support the mathematical modeling of the cell-cell dynamics in the tumor microenvironment. We performed bulk RNA-seq analysis and compared gene expression levels in the canine melanoma patient samples with those in the healthy canine samples. PCA analysis showed a clear distinction between tumor samples and healthy tissue samples.

Pathway enrichment analysis of DEGs showed that most of the significantly up- or down-regulated genes were related to immunological functions and immune related signaling pathways. Consistent with primary human melanoma, our analysis revealed upregulation of RAS pathway activity including increased PI3K expression and pathway activity. However, we also observed an unanticipated upregulation of the negative regulator, PTEN. Other classical human melanoma features observed in the canine melanoma samples include downregulation of E-cadherin, implicated in metastasis of human melanoma, loss of expression of tumor suppressor protein p53, and upregulation of positive cell cycle regulator cyclin D2.

We used immune cell deconvolution of gene expression levels to decide which immune cell types should be included in our smaller model. This resulted in a final reduced model that included melanoma, M1 and M2 macrophages, cytotoxic  T cells, and regulatory T cells. Parameter values in this reduced model were estimated from the literature or from clinical knowledge.

We performed global sensitivity analysis with Sobol and eFAST methods to analyze the model and determine which parameters were most-influential on the level of melanoma at 180 days. These influential parameters represent potential novel immunotherapy targets, and pathways that could be targeted in combination to achieve better results. The top six parameters, obtained from the same rank-ordering for both methods, are $K_M$, $r_M$, $\delta_M$, $\delta_C$, $r_C$, $\alpha_{2CM}$. These parameters mainly involve the death and growth of melanoma cells and cytotoxic T lymphocytes. These findings align with previous findings that boosting cytotoxic T lymphocytes improves a patient's outcome.[103] The parameter $\alpha_{2CM}$ is part of a pathway from M2 that decreases the effect of cytotoxic T lymphocytes, which may represent a novel target for therapeutic intervention.

Our top six most-influential parameters were not all globally structurally identifiable, but if we remove $\delta_M$, the remaining five parameters are globally identifiable. This means that, with adequate data, we would be able to estimate these five parameters uniquely. Fitting these parameters to time series data from individual patients would yield valid patient-specific “digital twins”. Digital twin models can be used for predicting patient outcomes and calculating regimens that yield optimal patient results.

The model we developed, once validated with time series data, could be expanded by adding compartments beyond the tumor microenvironment, such as a lymph node and blood compartments. Fitting our model to time series from multiple compartments would further increase confidence in the model components and parameter estimates.

One innovation in this work is the use of data from the comparative oncology model of canine melanoma patients to support our mathematical model building. This makes our model suitable for use with human patients with rare melanomas such as acral, mucosal, and uveal melanoma. Other innovations include the combination of the following methods into a single workflow: bulk RNA-seq data analysis and immune cell deconvolution comparing melanoma samples with healthy tissue to determine model components to reduce a large literature model, pathway analysis to inform tumor-immune dynamics, and sensitivity and identifiability analyses to determine parameters that are influential and can be estimated from adequate data. By sharing this workflow and code we have developed, we hope to support future work by others also modeling diseases.

# Appendix A. Details of the large initial model from the literature.

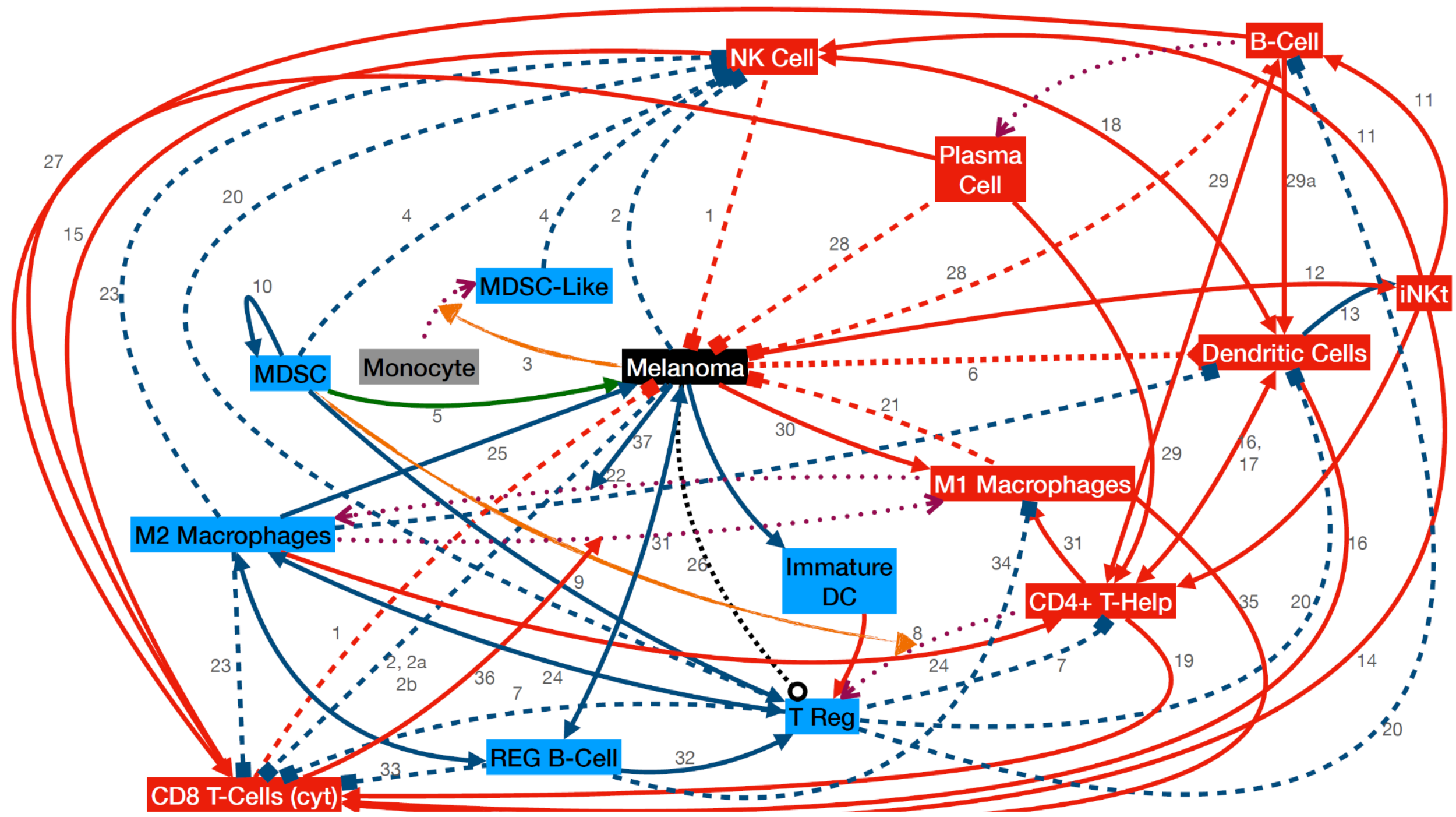

**Fig A1.** Large (initial, incomplete) model of melanoma-immune dynamics with 16 cell types included, based on literature information. Cell populations are color coded by anti-tumor (red) and pro-tumor (blue). Solid arrows represent a cell population's activation or induction. Dashed lines, which end in squares, represent inhibition of a cell population. The dotted black line from melanoma represents recruitment of T regs to the tumor microenvironment. Purple dotted lines represent transformations of cells. Orange arrows represent induction of cell changes.The green arrow indicates growth promotion. Any duplicate of pathway numbers means the biological dynamics of that pathway are the same, but the populations involved differ. See **Table A2** for detailed descriptions of each pathway.

We found in the literature multiple types of immune cells that were "cold", and did not attack the tumor. We included M2 macrophages, regulatory B cells (REG B-cells), immature dendritic cells (immature DC), regulatory T cells (TReg), myeloid-derived suppressor cells (MDSC), and MDSC-like cells. There were also immune cells that were "hot", and attempted to kill the tumor. We included natural killer (NK) cells, B lymphocyte (B) cells, plasma cells, dendritic cells, invariant natural killer T (iNKT) cells, M1 macrophages, CD4+ T-helper, CD8+ T cells. There was also another cell, the monocyte, which is neither antitumor or protumor as the other two classes of cells above. However, in cancer, monocyte cells may differentiate into many different types of immune cells such as tumor associated macrophages or dendritic cells. [115]

.

**Table A1.** Pathways for the large model of melanoma-immune dynamics.

| 1 | NK cells contribute to decreased tumor size by inducing apoptosis and releasing perforin and granzymes. | [116], [117] |
|---|---|---|
| 2 | Melanoma cells evade recognition by NK cells through mutations of MHC I and MHC II, and through release of VEGF, IL-8, and IL-10. | [116], [118] |
| 2a | Melanoma inhibits CD8+ T-cell function by inducing cytokines to exhaust them. Melanoma also expresses PDL1 and PDL2 which binds to PD1 on CD8+T cells to inhibit them. | [116] |
| 2b | Melanoma exhausts CD8+ T cells and induces apoptosis via Fas-FasL. | [118] |
| 3 | The melanoma TME releases PGE, which induces monocytes to differentiate to MDSC-like cells. | [119] |
| 4 | MDSC and MDSC-like cells release TGF-β, which leads to reduced CD247 on NK cells, and reduced IFN-γ, TNF-α and granzyme levels. They also release Arginase and NO which inhibit NK and T Cell antitumor functions. | [117], [116] |
| 5 | MDSCs directly release and indirectly promote the release of VEGF, FGF2, BV8, MMP's, and promote angiogenesis, which all cause melanoma growth. | [117] |
| 6 | Melanoma cells release IL8, IL10, TGF-β, and VEGF, inhibiting DC cells. | [118] |
| 7 | T-reg cells inhibit CTLs and T-helper cells via TGF-β, IL-10 and IDO overproduction. | [116], [118] |
| 8 | MDSCs help convert CD4+ T-helper cells into T-regs. | [117] |
| 9 | MDSCs help promote naive CD4+ T cells to differentiate to T-reg cells. | [117] |
| 10 | MDSCs release VEGF which increases recruitment of more MDSCs and contributes to immunosuppression. | [120] |
| 11 | iNKT activates NK and B cells once activated via production of large amounts of cytokines including IFN-γ. | [121] |
| 12 | Melanoma tumor cells release antigens that directly activate iNKT cells. | [121] |
| 13 | DCs present antigen from melanoma cells to iNKT. | [121], [118] |

| 14 | iNKt cells activate T cells to promote anti-tumor effects. | [121] |
|---|---|---|
| 15 | NK cells produce cytokines that help activate CD8+ T cells | [122] |
| 16 | DCs present antigens via MHC I to CD8+ T Cells and MHC II to CD4+ T Cells. | [118] |
| 17 | Th1 CD4+ T-helper cells can help DCs mature. | [118] |
| 18 | DCs present antigens to NK cells to activate them. | [118] |
| 19 | CD4+ T cells activate CD8+ T-cells, and release IFN-γ, TNF-α, and IL-2. | [123] |
| 20 | Tregs suppress anti-tumor functions of NK cells. | [122] |
| 21 | M1 macrophages present antigen and produce Th1 cytokines. | [118], [124] |
| 22 | Melanoma cells cause M1 macrophages to convert to M2 macrophages. | [118] |
| 23 | M2 macrophages inhibit NK cells’ and CD8 T cells’ anti tumor effects. | [118] |
| 24 | M2 macrophages promote Treg cell proliferation as well as Th2 cell proliferation. | [118] |
| 25 | M2 macrophages promote angiogenic effects that support tumor growth. | [118] |
| 26 | The melanoma TME’s chemokines attract CD4+ Tregs, which promote tumor growth and inhibit antitumor responses. | [118] |
| 27 | B and plasma cells activate CD8+ T-Cells via MHC I, and secretion of IFN-γ and IL-2. | [124] |
| 28 | B and plasma cells amplify antibodies against the melanoma tumor, specifically IgG1. | [124] |
| 29 | B and plasma cells activate CD4 T-Cells via MHC II, and secretion of IFN-γ, TNF-α and IL-2 | [124] |
| 29a | B cells induce maturation of dendritic cells. | [125] |
| 30 | Melanoma tumor cells release IFN-γ which upregulates M1 macrophages. | [124] |
| 31 | Melanoma tumor cells secrete certain tumor antigens which prompt B-cells to become Reg B-Cells. Reg B-cells then promote secretion of IL-10, TGF-β, and class switching to IgG4, a protumorigenic antibody. | [126] |
| 32 | Reg B cells stimulate Tregs both through cell-cell contact and through production IL-10. | [127],[128] |
| 33 | Reg B cells inhibit CD8+ T cells via TCR, PD/PD-L1 interactions, and IL-10. | [129] |
| 34 | Reg B cells inhibit M1 macrophages by secreting IL-10. | [130] |

## Appendix B. GO terms for original analysis.

| Category | Term | Count | p-value |
|---|---|---|---|
| BP | Immune system process | 216 | 9.5e-147 |
| BP | Positive regulation of biological process | 211 | 4.7e-62 |
| BP | Immune response | 172 | 2.6e-127 |
| BP | Response to external stimulus | 159 | 6.3e-83 |
| BP | Response to stress | 158 | 5.9e-58 |
| CC | Cell periphery | 151 | 4.0e-22 |
| CC | Plasma membrane | 143 | 5.4e-22 |
| CC | Extracellular region | 102 | 7.6e-36 |
| CC | Cell surface | 85 | 4.4e-56 |
| CC | Extracellular space | 79 | 4.8e-32 |
| MF | Signaling receptor binding | 104 | 3.5e-52 |
| MF | Signaling receptor activity | 83 | 6.8e-21 |
| MF | Molecular transducer activity | 83 | 6.8e-21 |
| MF | Transmembrane signaling receptor activity | 69 | 3.4e-15 |
| MF | Cytokine receptor binding | 53 | 2.7e-47 |

**Table B1.** The top 5 enriched GO terms which show up-regulated pathways from Biological Processes (BP), Cell Components (CC), and Molecular Functions (MF) are selected.

| Category | Term | Count | p-value |
|---|---|---|---|
| BP | Signaling | 104 | 1.0e-32 |
| BP | Positive regulation of biological process | 96 | 1.0e-31 |
| BP | Cell surface receptor signaling pathway | 83 | 1.7e-46 |
| BP | Regulation of response to stimulus | 80 | 1.1e-33 |
| BP | Positive regulation of metabolic process | 80 | 5.2e-33 |
| CC | Cell periphery | 63 | 5.5e-09 |
| CC | Plasma membrane | 60 | 5.2e-09 |
| CC | Cytosol | 41 | 8.4e-06 |
| CC | Cell surface | 29 | 1.8e-15 |
| CC | Intrinsic component of plasma membrane | 26 | 3.1e-08 |
| MF | Identical protein binding | 47 | 1.3e-19 |
| MF | Enzyme binding | 45 | 8.8e-17 |
| MF | Signaling receptor binding | 36 | 1.4e-14 |
| MF | Kinase activity | 25 | 3.6e-10 |
| MF | Phosphotransferase activity, alcohol group as acceptor | 25 | 5.4e-11 |

**Table B2.** The top 5 enriched GO terms which show down-regulated pathways from Biological

# Appendix C. Gene expression and immune deconvolution for additional data from canine lymph nodes.

### Lymph Nodes Analysis

Initially, we had pre-vaccine canine melanoma and healthy canine lymph node (LN) tissue samples. We started our study by performing immune cell deconvolution on cancerous lymph node samples and healthy lymph node samples. The lymphatic system is used to remove waste and cellular debris from the body, and is an essential component of the immune system and other bodily functions. Lymph nodes are various sites within the lymphatic system where this debris and drainage can be processed or destroyed by immune cells. Primary tumors have the ability to evade immune responses and spread throughout the host. This can be accomplished via hosts' lymphatic systems. Despite the fact that the LN is a proximal site for mounting immune defense against local tumors, the LN is the most-common site for early tumor metastasis.[47,48] Lymphatic vessels near the tumor can act as a conduit for malignant cells from the primary tumor to lymph nodes. From there, malignant cells can disseminate into the systemic circulation and to other organs, causing metastasis.[47] As such, LN metastasis is an important outcome predictor for certain cancers.[47]

The results of the immune cell deconvolution show CD8+ T cells are one of the key immune cells, however, there were no significant changes between lymph node tumor samples and lymph node healthy samples. We then compared primary tumor samples and healthy samples. The results show the top four significantly different immune cells between samples were T regulatory cells ($p = 0.0002$), M2 macrophages ($p = 0.0028$), CD8+ T cells ($p = 0.0218$), and M1 macrophages ($p = 0.0558$).

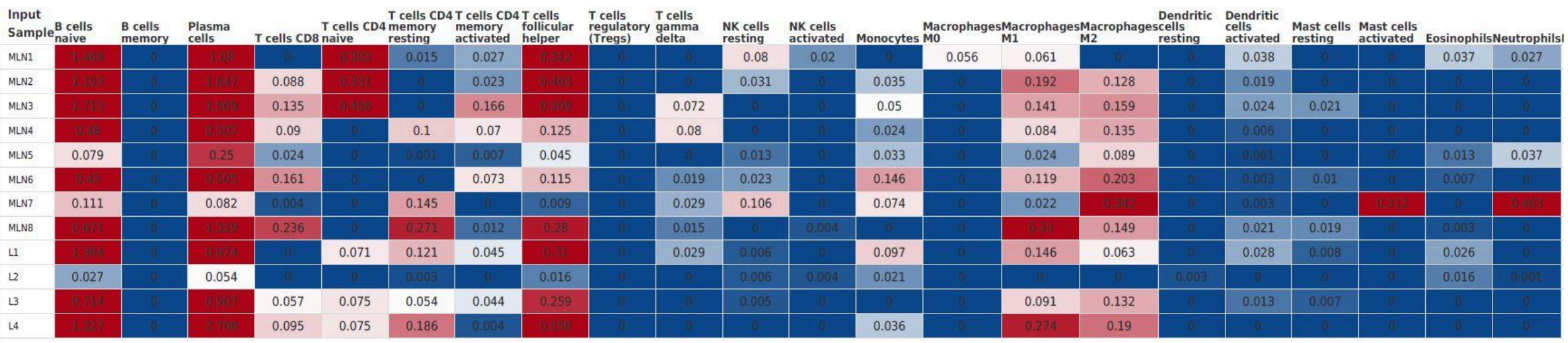

| Input Sample | B cells naive | B cells memory | Plasma cells | T cells CD8 | T cells CD4 naive | T cells CD4 memory resting | T cells CD4 memory activated | T cells follicular helper | T cells regulatory (Tregs) | T cells gamma delta | NK cells resting | NK cells activated | Monocytes | Macrophages M0 | Macrophages M1 | Macrophages M2 | Dendritic cells resting | Dendritic cells activated | Mast cells resting | Mast cells activated | Eosinophils | Neutrophils |
| --- | --- | --- | --- | --- | --- | --- | --- | --- | --- | --- | --- | --- | --- | --- | --- | --- | --- | --- | --- | --- | --- | --- |
| MLN1 | 1.468 | 0 | 1.08 | 0 | 0.309 | 0.015 | 0.027 | 0.312 | 0 | 0 | 0.08 | 0.02 | 0 | 0.056 | 0.061 | 0 | 0 | 0.038 | 0 | 0 | 0.037 | 0.027 |
| MLN2 | 1.193 | 0 | 1.647 | 0.088 | 0.331 | 0 | 0.023 | 0.463 | 0 | 0 | 0.031 | 0 | 0.035 | 0 | 0.192 | 0.128 | 0 | 0.019 | 0 | 0 | 0 | 0 |
| MLN3 | 1.713 | 0 | 1.509 | 0.135 | 0.458 | 0 | 0.166 | 0.309 | 0 | 0.072 | 0 | 0 | 0.05 | 0 | 0.141 | 0.159 | 0 | 0.024 | 0.021 | 0 | 0 | 0 |
| MLN4 | 0.48 | 0 | 0.507 | 0.09 | 0 | 0.1 | 0.07 | 0.125 | 0 | 0.08 | 0 | 0 | 0.024 | 0 | 0.084 | 0.135 | 0 | 0.006 | 0 | 0 | 0 | 0 |
| MLN5 | 0.079 | 0 | 0.25 | 0.024 | 0 | 0.001 | 0.007 | 0.045 | 0 | 0 | 0.013 | 0 | 0.033 | 0 | 0.024 | 0.089 | 0 | 0.001 | 0 | 0 | 0.013 | 0.037 |
| MLN6 | 0.42 | 0 | 0.505 | 0.161 | 0 | 0 | 0.073 | 0.115 | 0 | 0.019 | 0.023 | 0 | 0.146 | 0 | 0.119 | 0.203 | 0 | 0.003 | 0.01 | 0 | 0.007 | 0 |
| MLN7 | 0.111 | 0 | 0.082 | 0.004 | 0 | 0.145 | 0 | 0.009 | 0 | 0.029 | 0.106 | 0 | 0.074 | 0 | 0.022 | 0.342 | 0 | 0.003 | 0 | 0.312 | 0 | 0.463 |
| MLN8 | 0.671 | 0 | 1.329 | 0.236 | 0 | 0.271 | 0.012 | 0.28 | 0 | 0.015 | 0 | 0.004 | 0 | 0 | 0.33 | 0.149 | 0 | 0.021 | 0.019 | 0 | 0.003 | 0 |
| L1 | 1.364 | 0 | 0.974 | 0 | 0.071 | 0.121 | 0.045 | 0.31 | 0 | 0.029 | 0.006 | 0 | 0.097 | 0 | 0.146 | 0.063 | 0 | 0.028 | 0.008 | 0 | 0.026 | 0 |
| L2 | 0.027 | 0 | 0.054 | 0 | 0 | 0.003 | 0 | 0.016 | 0 | 0 | 0.006 | 0.004 | 0.021 | 0 | 0 | 0 | 0.003 | 0 | 0 | 0 | 0.016 | 0.001 |
| L3 | 0.714 | 0 | 0.907 | 0.057 | 0.075 | 0.054 | 0.044 | 0.259 | 0 | 0 | 0.005 | 0 | 0 | 0 | 0.091 | 0.132 | 0 | 0.013 | 0.007 | 0 | 0 | 0 |
| L4 | 1.527 | 0 | 2.708 | 0.095 | 0.075 | 0.186 | 0.004 | 0.558 | 0 | 0 | 0 | 0 | 0.036 | 0 | 0.274 | 0.19 | 0 | 0 | 0 | 0 | 0 | 0 |

**Fig C1.** Heat map of estimated proportions of 22 immune cells for twelve canine samples (8 lymph node tumor samples, 4 lymph node healthy samples). We used CIBERSORTx to generate the figure.

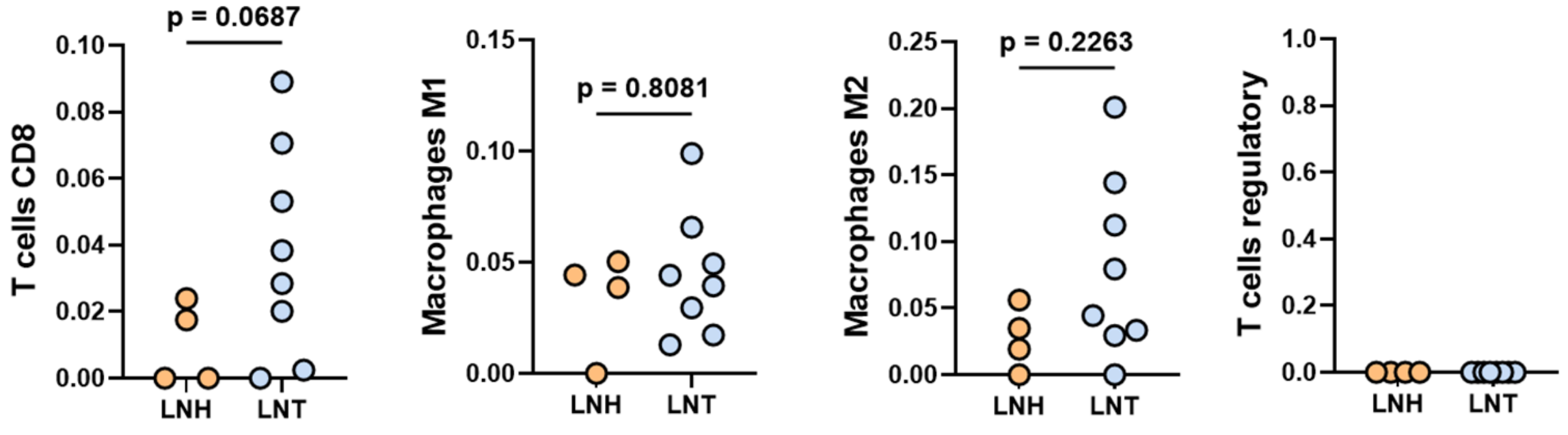

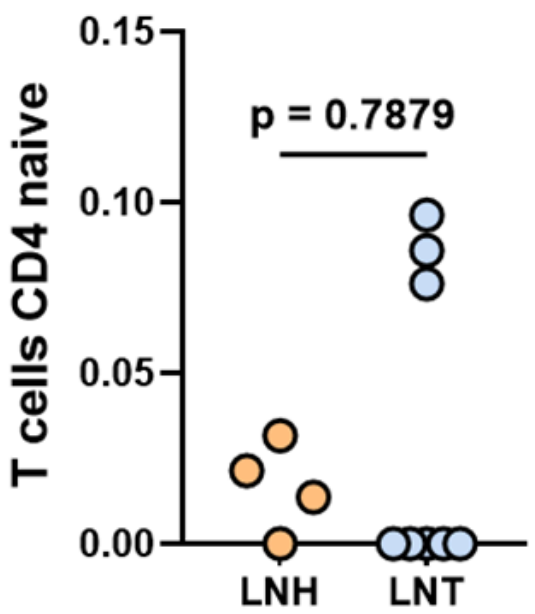

**Fig C2.** Immune cell deconvolution was conducted using CIBERSORTx algorithm. Five immune cell types are shown. Figures for cell levels were produced with PRISM. All p-values were calculated using the Mann-Whitney U test.

The lack of significant differences between these two groups may be due to the healthy samples being in the draining region from the tumor, leading to a similar phenotype.

## Lymph node tumor samples vs primary tumor samples

We analyzed the tumor sample versus the healthy sample and the lymph node tumor samples versus the lymph node healthy samples. In this section we analyze the bulk RNA-seq data of the tumor samples and lymph node tumor samples. We believe these analyses offer valuable insights about tumor evolution and pathways during metastasis. Moreover, we may learn more about adaptation of tumor cells in different sites. In this study, we have tumor samples and their paired samples from lymph nodes tumors. Immune cell deconvolution showed M1 macrophages were statistically significantly different between samples ($p = 0.0447$), however, because all samples were from tumors, other immune cells such as M2 macrophages or CD8+ T cells did not show significant differences.

We calculated the average value for each of the 22 immune cells whose proportional values derived based on absolute value resulted from CIBERSORTx. The calculations are shown in **Table B3** for primary and healthy samples alongside cancerous and healthy lymph node samples.

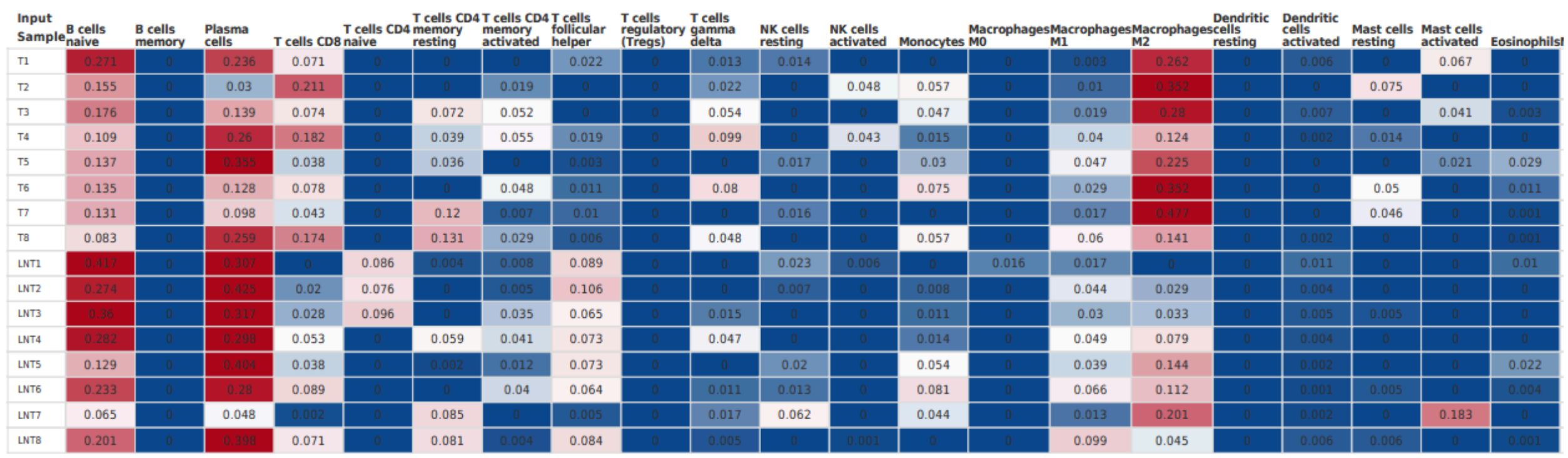

| Input Sample | B cells naive | B cells memory | Plasma cells | T cells CD8 | T cells CD4 naive | T cells CD4 memory resting | T cells CD4 memory activated | T cells follicular helper | T cells regulatory (Tregs) | T cells gamma delta | NK cells resting | NK cells activated | Monocytes | Macrophages M0 | Macrophages M1 | Macrophages M2 | Dendritic cells resting | Dendritic cells activated | Mast cells resting | Mast cells activated | Eosinophils |
|---|---|---|---|---|---|---|---|---|---|---|---|---|---|---|---|---|---|---|---|---|---|
| T1 | 0.271 | 0 | 0.236 | 0.071 | 0 | 0 | 0 | 0.022 | 0 | 0.013 | 0.014 | 0 | 0 | 0 | 0.003 | 0.262 | 0 | 0.006 | 0 | 0.067 | 0 |
| T2 | 0.155 | 0 | 0.03 | 0.211 | 0 | 0 | 0.019 | 0 | 0 | 0.022 | 0 | 0.048 | 0.057 | 0 | 0.01 | 0.352 | 0 | 0 | 0.075 | 0 | 0 |
| T3 | 0.176 | 0 | 0.139 | 0.074 | 0 | 0.072 | 0.052 | 0 | 0 | 0.054 | 0 | 0 | 0.047 | 0 | 0.019 | 0.28 | 0 | 0.007 | 0 | 0.041 | 0.003 |
| T4 | 0.109 | 0 | 0.26 | 0.182 | 0 | 0.039 | 0.055 | 0.019 | 0 | 0.099 | 0 | 0.043 | 0.015 | 0 | 0.04 | 0.124 | 0 | 0.002 | 0.014 | 0 | 0 |
| T5 | 0.137 | 0 | 0.355 | 0.038 | 0 | 0.036 | 0 | 0.003 | 0 | 0 | 0.017 | 0 | 0.03 | 0 | 0.047 | 0.225 | 0 | 0 | 0 | 0.021 | 0.029 |
| T6 | 0.135 | 0 | 0.128 | 0.078 | 0 | 0 | 0.048 | 0.011 | 0 | 0.08 | 0 | 0 | 0.075 | 0 | 0.029 | 0.352 | 0 | 0 | 0.05 | 0 | 0.011 |
| T7 | 0.131 | 0 | 0.098 | 0.043 | 0 | 0.12 | 0.007 | 0.01 | 0 | 0 | 0.016 | 0 | 0 | 0 | 0.017 | 0.477 | 0 | 0 | 0.046 | 0 | 0.001 |
| T8 | 0.083 | 0 | 0.259 | 0.174 | 0 | 0.131 | 0.029 | 0.006 | 0 | 0.048 | 0 | 0 | 0.057 | 0 | 0.06 | 0.141 | 0 | 0.002 | 0 | 0 | 0.001 |
| LNT1 | 0.417 | 0 | 0.307 | 0 | 0.086 | 0.004 | 0.008 | 0.089 | 0 | 0 | 0.023 | 0.006 | 0 | 0.016 | 0.017 | 0 | 0 | 0.011 | 0 | 0 | 0.01 |
| LNT2 | 0.274 | 0 | 0.425 | 0.02 | 0.076 | 0 | 0.005 | 0.106 | 0 | 0 | 0.007 | 0 | 0.008 | 0 | 0.044 | 0.029 | 0 | 0.004 | 0 | 0 | 0 |
| LNT3 | 0.36 | 0 | 0.317 | 0.028 | 0.096 | 0 | 0.035 | 0.065 | 0 | 0.015 | 0 | 0 | 0.011 | 0 | 0.03 | 0.033 | 0 | 0.005 | 0.005 | 0 | 0 |
| LNT4 | 0.282 | 0 | 0.298 | 0.053 | 0 | 0.059 | 0.041 | 0.073 | 0 | 0.047 | 0 | 0 | 0.014 | 0 | 0.049 | 0.079 | 0 | 0.004 | 0 | 0 | 0 |
| LNT5 | 0.129 | 0 | 0.404 | 0.038 | 0 | 0.002 | 0.012 | 0.073 | 0 | 0 | 0.02 | 0 | 0.054 | 0 | 0.039 | 0.144 | 0 | 0.002 | 0 | 0 | 0.022 |
| LNT6 | 0.233 | 0 | 0.28 | 0.089 | 0 | 0 | 0.04 | 0.064 | 0 | 0.011 | 0.013 | 0 | 0.081 | 0 | 0.066 | 0.112 | 0 | 0.001 | 0.005 | 0 | 0.004 |
| LNT7 | 0.065 | 0 | 0.048 | 0.002 | 0 | 0.085 | 0 | 0.005 | 0 | 0.017 | 0.062 | 0 | 0.044 | 0 | 0.013 | 0.201 | 0 | 0.002 | 0 | 0.183 | 0 |
| LNT8 | 0.201 | 0 | 0.398 | 0.071 | 0 | 0.081 | 0.004 | 0.084 | 0 | 0.005 | 0 | 0.001 | 0 | 0 | 0.099 | 0.045 | 0 | 0.006 | 0.006 | 0 | 0.001 |

**Fig C3.** Heat map of estimated proportions of 22 immune cells for twelve canine samples (8 lymph node tumor samples, 8 primary tumor samples). CD8+ T cells, M1 macrophages, M2 macrophages, and T regulatory cells show the largest difference between groups. We used CIBERSORTx to generate the figure.

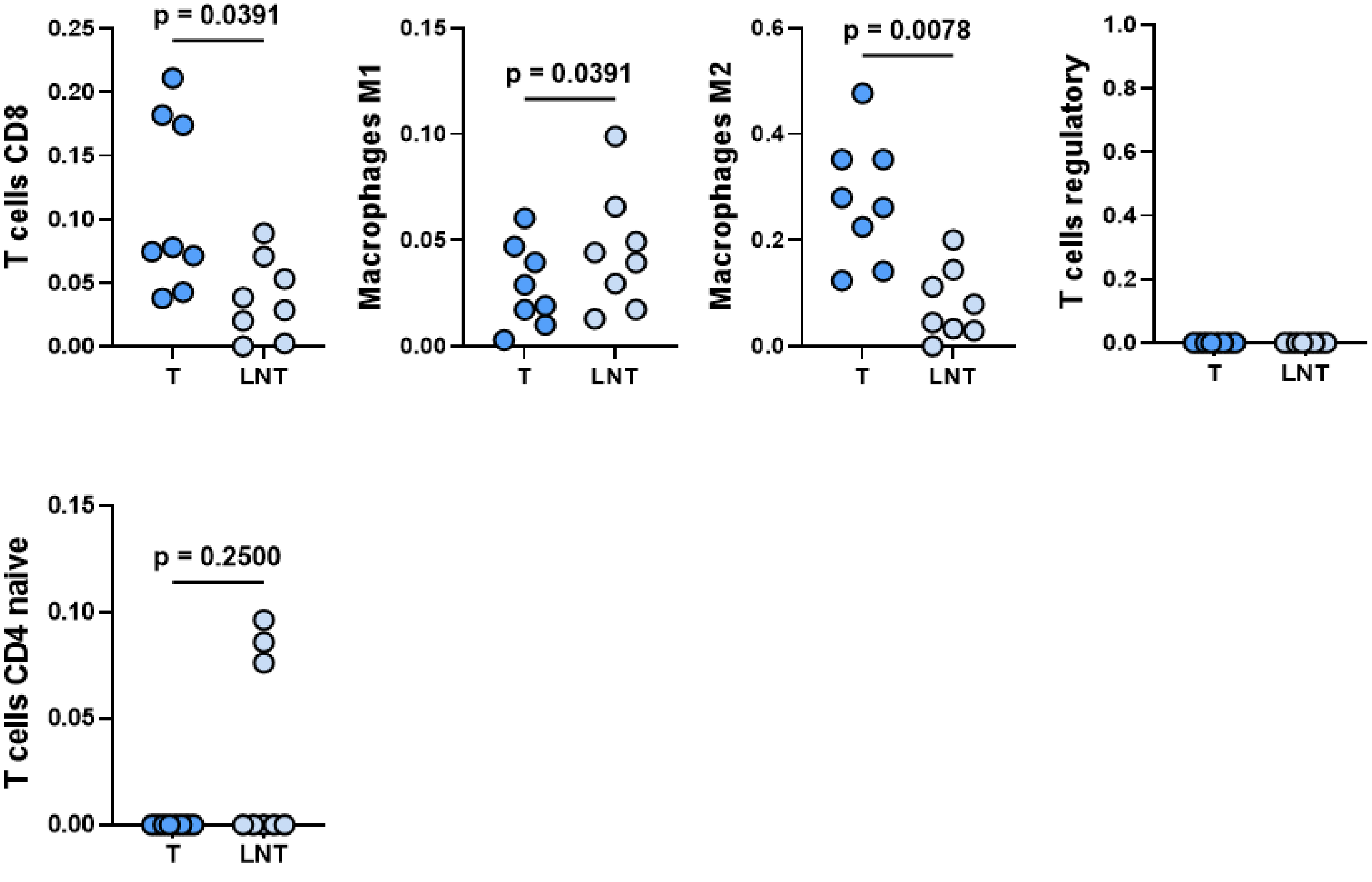

**Fig C4.** Immune cell deconvolutions were conducted using the CIBERSORTx algorithm. The five immune cell types' proportions between samples are shown. All p-values were calculated by paired t-tests. Figures were generated with PRISM.

## RNA-seq Analysis

### Lymph nodes

Initially, we collected four healthy lymph node samples, however, one of them was excluded in our analysis due to a flag in a control linearity test.

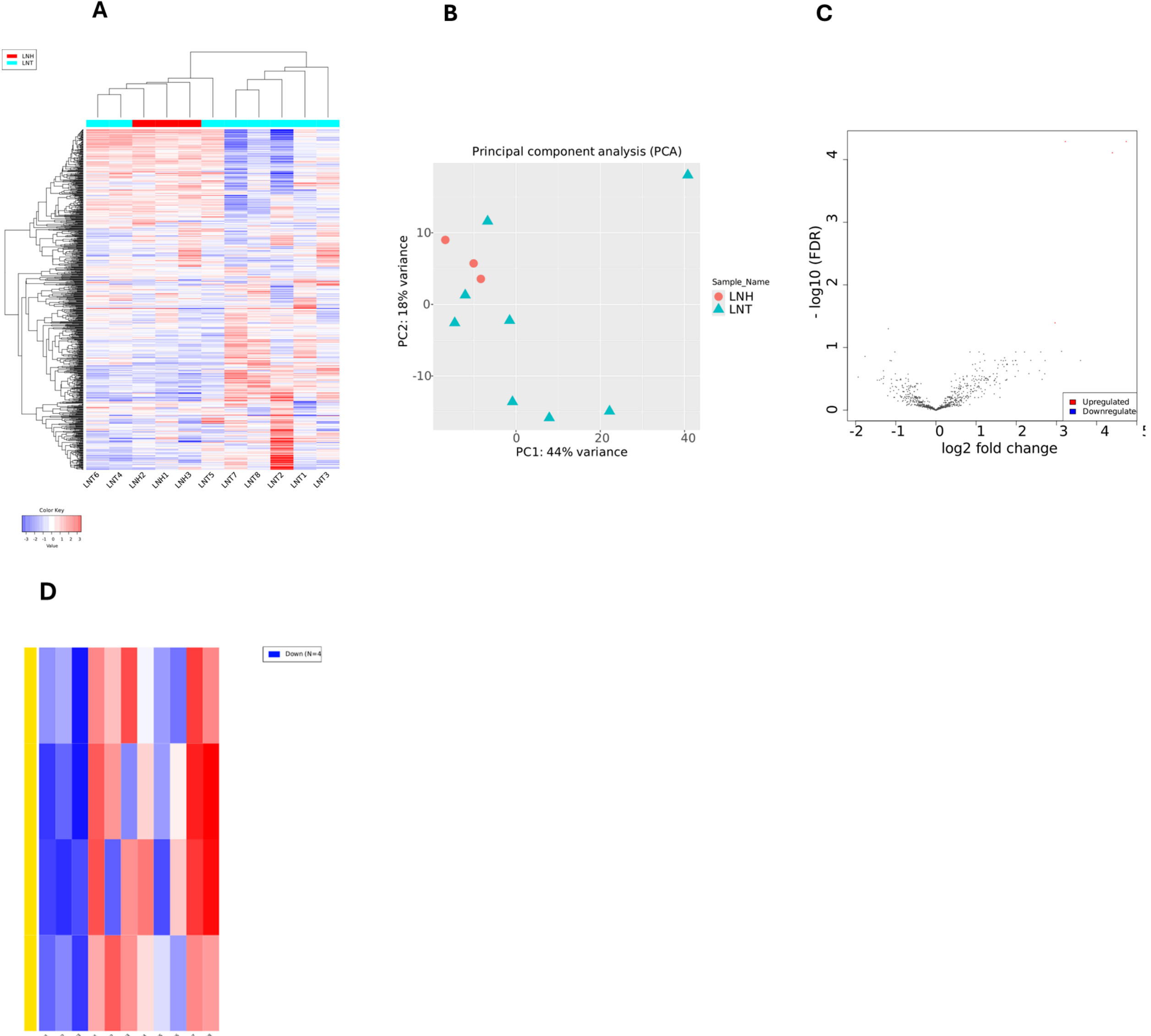

**Fig C5.** Hierarchical clustering heat map (A) and principal component analysis (B) examine the results of DEGs for each up and down comparison (C) Enrichment pathways in DEGs for the selected comparison (D) for lymph nodes.

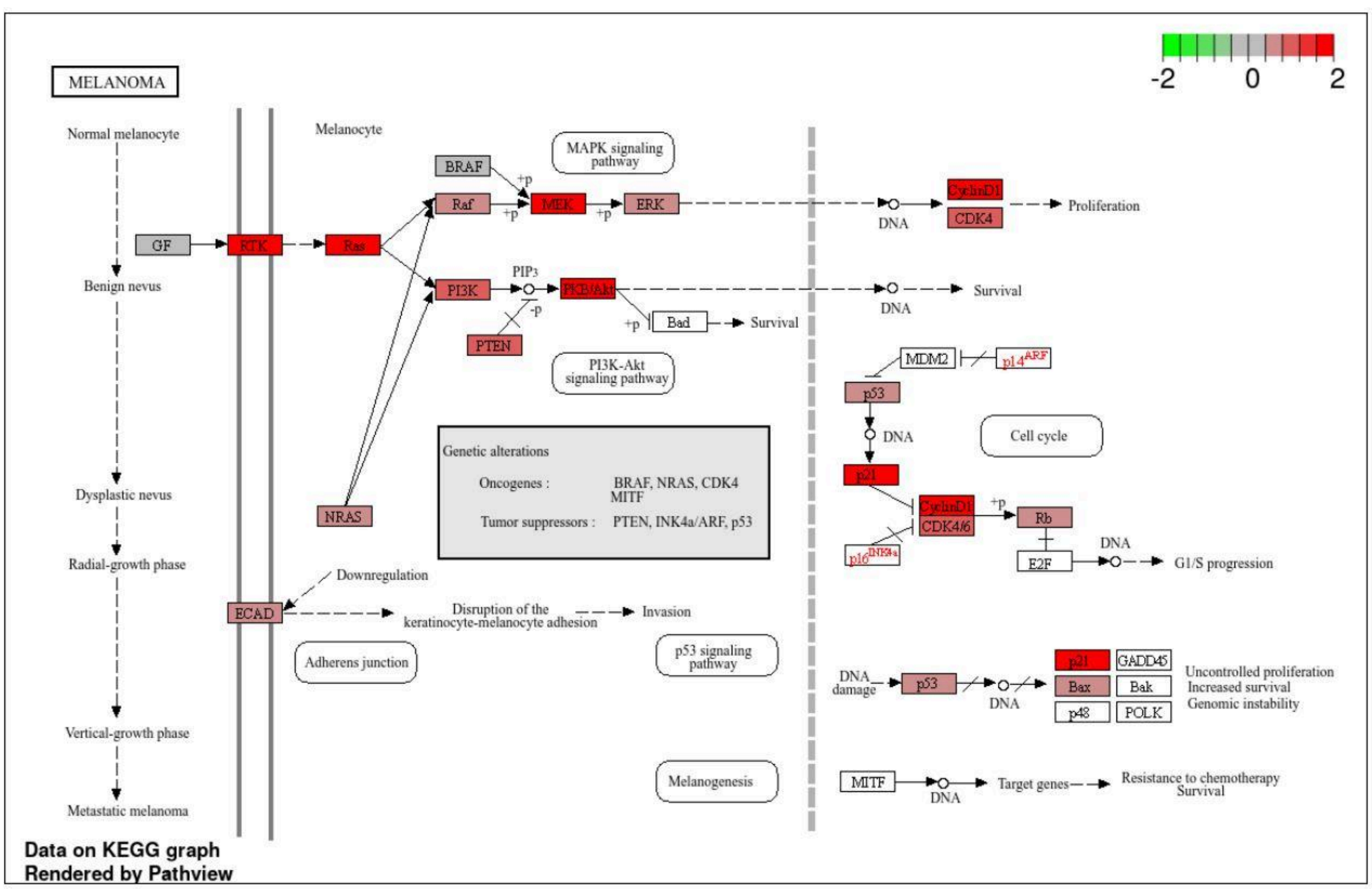

**Fig C6.** Visualization of pathway enrichment analysis of LN samples from canine melanoma patients vs. LN samples from healthy canine patients. Expression profiles of cell-cycle related genes visualized on a melanoma KEGG signaling pathway diagram using Pathview. Red and green labels indicate genes upregulated or downregulated, respectively.

**Lymph node tumors vs primary tumor**

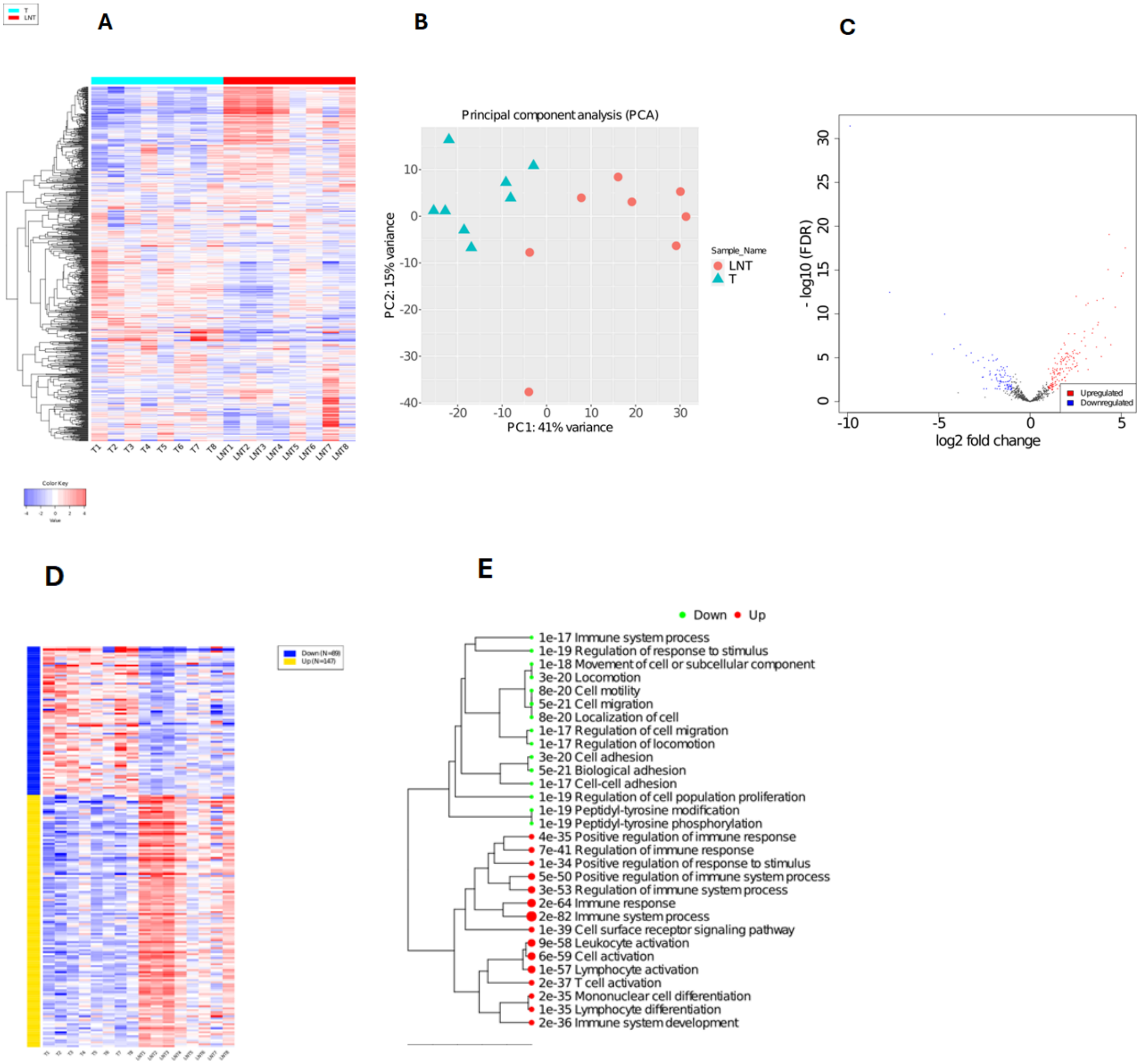

**Fig C7.** Exploration of DEGs. (A) Hierarchical clustering heat map. (B) Principal component analysis. (C) Enrichment pathways in DEGs for the selected comparison. (D) Heat map of DEGs for the lymph node tumor samples vs primary tumors. (E)

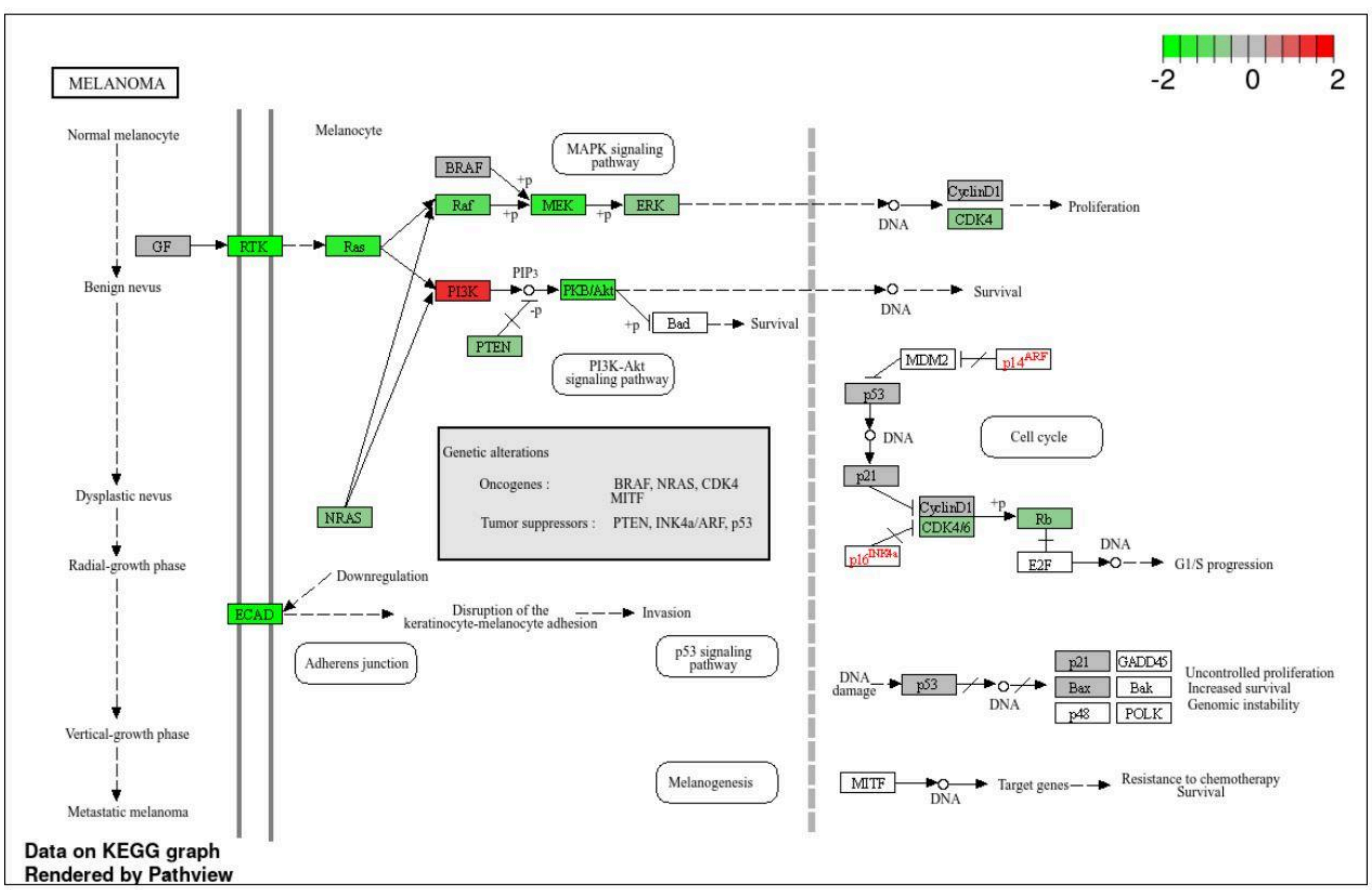

**Fig C8.** Visualization of pathway enrichment analysis of LN samples from canine melanoma patients vs. primary tumor samples from canine melanoma patients. Expression profiles of cell-cycle related genes visualized on a melanoma KEGG signaling pathway diagram using Pathview. Red and green labels indicate genes upregulated or downregulated, respectively.

## Appendix D. Additional sensitivity analysis details.

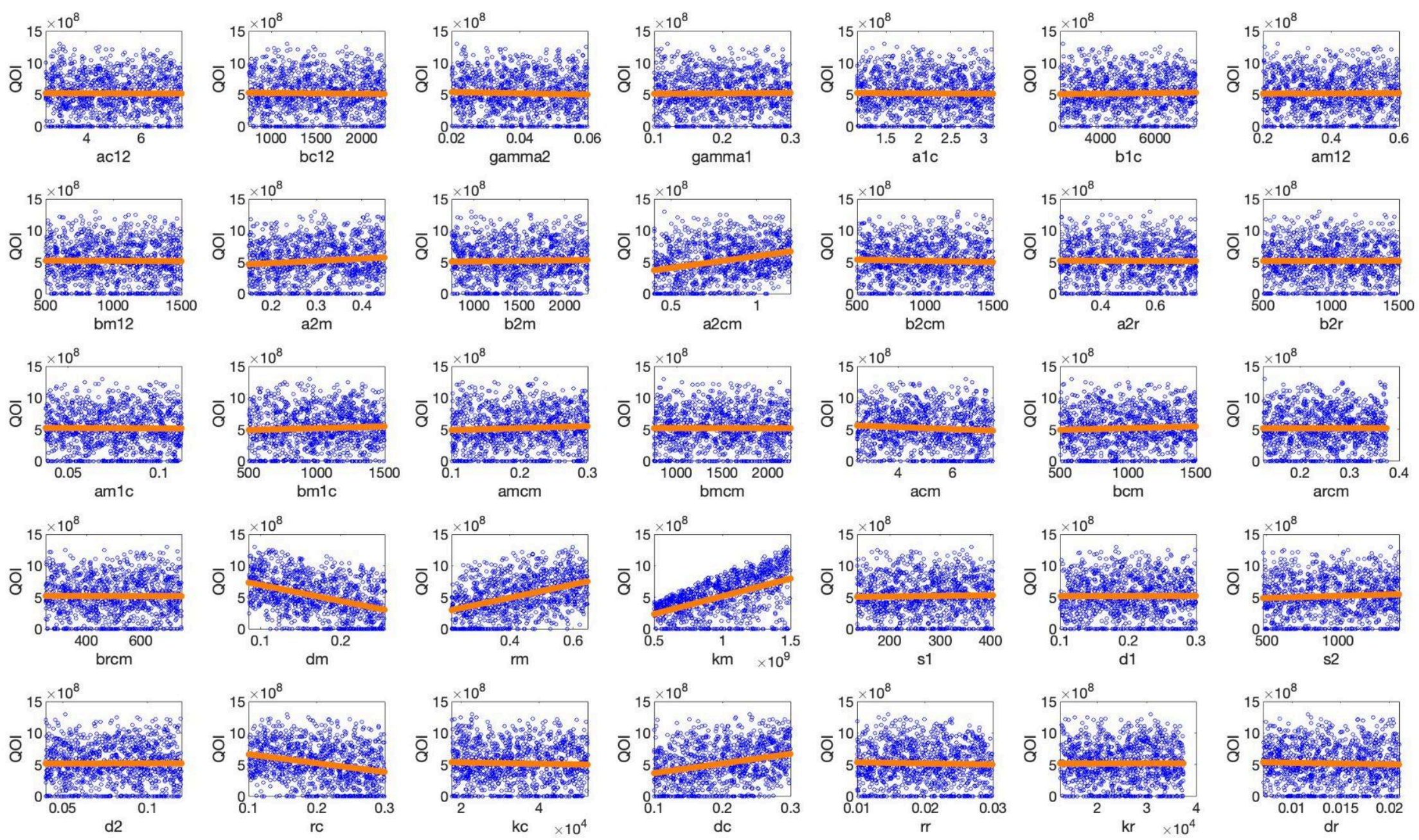

**Fig D1**. Individual scatterplots of each parameter vs QOI value. The blue points are 1,000 randomly selected QOI values from the 6,475,000 QOI values that resulted from performing the original Sobol sensitivity analysis, and the parameter values that produced them. Overlaid onto each scatter plot is a least-squares regression line in orange. As expected, influential parameters have regression lines with slopes visually different than 0. Statistical significance of these slopes is summarized in Table D1.

| Parameter | p-value |
|---|---|
| $K_M$ | 5.2707e-71 |
| $r_M$ | 1.5575e-43 |
| $\delta_M$ | 3.1732e-41 |
| $\delta_C$ | 4.8487e-20 |
| $\alpha_{2CM}$ | 3.1359e-19 |
| $r_C$ | 4.0569e-17 |
| $\alpha_{2M}$ | 0.0013 |
| $\alpha_{CM}$ | 0.0153 |
| $\alpha_{MCM}$ | 0.0652 |

| $s_2$ | 0.0695 |
|---|---|
| $\beta_{M1C}$ | 0.1107 |
| $\beta_{CM}$ | 0.1111 |
| $\gamma_2$ | 0.2501 |
| $\beta_{2CM}$ | 0.2533 |
| $K_C$ | 0.2585 |
| $\delta_r$ | 0.2740 |
| $r_r$ | 0.3088 |
| $\beta_{2M}$ | 0.4284 |
| $\beta_{1C}$ | 0.4616 |
| $s_1$ | 0.4711 |
| $\beta_{C12}$ | 0.5924 |
| $\alpha_{1C}$ | 0.6904 |
| $\gamma_1$ | 0.7264 |
| $\beta_{M12}$ | 0.7722 |
| $\alpha_{M1C}$ | 0.7856 |
| $\alpha_{M12}$ | 0.8156 |
| $\alpha_{2R}$ | 0.8268 |
| $\alpha_{C12}$ | 0.8468 |
| $\alpha_{RCM}$ | 0.9107 |
| $\delta_1$ | 0.9146 |
| $\beta_{RCM}$ | 0.9273 |
| $\delta_2$ | 0.9345 |
| $\beta_{2R}$ | 0.9359 |

| $K_R$ | 0.9477 |
| --- | --- |
| $\beta_{MCM}$ | 0.9920 |

**Table D1.** Linear regression was performed to determine the significance of the relationship between each parameter and the QOI. The same subset of 1,000 points from **Fig D1** were used in MATLAB's lmfit function and the p-values of the slopes were extracted and sorted by ascending value. Notably, the order of the 6 most-influential parameters' p-values matches the order of their total sensitivity indices; furthermore, there is a sizable increase in magnitude in p-value after the 6 most-influential parameters. This aligns with our decision to use only the top 6 most-influential parameters in our analysis.